\documentclass[useAMS,usenatbib,usegraphicx]{mn2e}

\usepackage{amsmath,amssymb}
\usepackage[T1]{fontenc}
\usepackage{times}

\newcommand{\kms}{\ensuremath{\,\mathrm{km}\,\mathrm{s}^{-1}}}

\newcommand{\pcmc}{\ensuremath{\,\mathrm{cm}^{-3}}}
\newcommand{\pcms}{\ensuremath{\,\mathrm{cm}^{-2}}}
\newcommand{\pcmsps}{\ensuremath{\,\mathrm{cm}^{-2}\,\mathrm{s}^{-1}}}
\newcommand{\pcmcps}{\ensuremath{\,\mathrm{cm}^{-3}\,\mathrm{s}^{-1}}}
\newcommand{\hii}{H~\textsc{II}}

\newcommand{\msun}{\ensuremath{\mathrm{M}_{\odot}}}

\begin{document}

\title[Formation of elephant trunks in \hii\ regions]
{Dynamical models for the formation of elephant trunks in \hii~
  regions}
 \author[J.\ Mackey and A.J.\ Lim]
 	{Jonathan Mackey$^1$\thanks{email: \texttt{jmackey@cp.dias.ie}} and 
          Andrew J. Lim$^1$\thanks{
            Current address: Cambridge Silicon Radio Ltd., Churchill
            House, Cambridge Business Park, Cowley Road, Cambridge,
            CB4 0WZ, UK. (email: \texttt{Andrew.Lim@csr.com})}\\
  	$^1$Dublin Institute for Advanced Studies, 31 Fitzwilliam Place,
	Dublin 2, Ireland}
\date{Accepted 2009 December 7.  Received 2009 November 30; in original form 2009 September 25. \\
  The definitive version is available at \texttt{http://www.blackwell-synergy.com}
}
\maketitle

\begin{abstract}
  The formation of pillars of dense gas at the boundaries of \hii\
  regions is investigated with hydrodynamical numerical simulations
  including ionising radiation from a point source.  We show that
  shadowing of ionising radiation by an inhomogeneous density field is
  capable of forming so-called \emph{elephant trunks} (pillars of
  dense gas as in e.g.~M16) without the assistance of self-gravity, or
  of ionisation front and cooling instabilities.  A large simulation
  of a density field containing randomly generated clumps of gas is
  shown to naturally generate elephant trunks with certain clump
  configurations.  These configurations are simulated in isolation and
  analysed in detail to show the formation mechanism and determine
  possible observational signatures.  Pillars formed by the shadowing
  mechanism are shown to have rather different velocity profiles
  depending on the initial gas configuration, but asymmetries mean
  that the profiles also vary significantly with perspective, limiting
  their ability to discriminate between formation scenarios.  Neutral
  and molecular gas cooling are shown to have a strong effect on these
  results.
\end{abstract}


\section{Introduction}
Among the most striking features of \hii\ regions are the long columns
of neutral gas pointing towards the central star, known variously as
elephant trunks (ETs), pillars and fingers.  The most famous examples
are the `pillars of creation' in the Eagle Nebula (M16), observed with
HST by~\citet{HesScoSanEA96}.  The HST images of NGC 3372 (the Carina
Nebula) clearly show a number of ETs, as do HST images of NGC
3603~\citep{BraGreChuEA00}.  The Elephant Trunk Nebula in IC 1396,
observed by e.g.~\citet{ReaRhoYouEA04} is another typical example.
\citet*{CarGahKri03} show ETs in the Rosette Nebula (NGC 2237-2246),
IC 1805, and N81.

The M16 pillars have been well studied observationally.  They are
towards the more massive end of the scale of observed ETs but are not
extreme, so we have used their properties as a benchmark for our
simulation results.  \citet{HesScoSanEA96} studied the surface of the
M16 pillars in detail with HST, finding that the interface between the
pillars and the \hii\ region is well explained by a thin ionisation
front (I-front) with a strong photo-evaporation flow into the lower
density ionised gas.  Due to the large opacity of the molecular gas
they could say little about the interior of the pillars.  Molecular CO
maps were obtained by~\citet{Pou98}, who showed that the pillars are
mostly molecular, that the internal and external pressures are
comparable, and that their photo-evaporation times are about
$2\times10^7$ years.  He also found strong velocity gradients in the
molecular emission, but it is not clear if this indicates a coherent
flow generated by pillar formation.  \citet{WhiNelHolEA99}, using
infrared, millimetre and radio observations, found the pillars to have
cold cores ($\sim 20\,$K), with surrounding warm ($\sim60\,$K) gas,
and an outer `hot' shell (at $\sim250-320\,$K).  The clumps at the
heads of the pillars appeared to have typical densities
$2\times10^5\pcmc$, while the trunks exhibited lower mean densities of
$3-5\times10^4\pcmc$. These authors also found that the internal
pressure of the heads of the pillars is a factor of $\sim2$ lower than
the adjacent ionised gas pressure.  They interpreted this as evidence
that the heads of the pillars are in the early stages of radiatively
driven implosion~\citep{Ber89}, and may therefore be $<100\,$kyr old.
Subsequent dynamical models by~\citet*{WilWarWhi01} suggest alternate
scenarios in which the pillars could be $200-400\,$kyr old.

Despite detailed observations, and a long history of theoretical
models, there is as yet no clear consensus as to how such features as
ETs arise, their lifetimes, or their end states.  It has been
suggested that instabilities in I-fronts can generate similar
structures or, alternatively, that pre-existing inhomogeneities in the
interstellar medium (ISM) will create shadowed regions behind dense
clumps where gas can accumulate.  The question of which of these
scenarios actually forms ETs was discussed at least as far back
as~\citet{Kah58}.  Given that the ISM is clumpy, and that
instabilities are present under certain conditions, it is likely that
both processes contribute to some degree.  Thus the difficulty from a
theoretical perspective is that there are a number of potentially
viable mechanisms to form pillars, and it is difficult to
observationally distinguish between them.  The instability model was
proposed by~\citet{Fri54} for Rayleigh-Taylor instabilities, and
subsequently developed for I-front instabilities.  These instabilities
have been studied by many authors but are not the subject of this
paper, so we refer the reader to recent work by e.g.\
\citet{Wil02,MizKanPouEA06,WhaNor08} and references therein.  We
concentrate here on modelling the effects of the shadowing of dense
clumps on the dynamics of an expanding I-front.  ~\citet{LimMel03}
found that a partially shadowed clump in an ionising radiation field
was pushed further into the shadowed region, and a denser and
longer-lived tail arose compared to that of a single isolated clump.
This was in the context of Earth-mass clumps in planetary
nebulae~\citep{ODelHan96} which have lifetimes of a few thousand
years.  For more massive clumps, the longer lifetimes may allow much
denser tails to develop in the shadowed regions, possibly to the
extent that they would be observed as pillars.  This was also
suggested by~\citet{PouKanRyuEA07}.

General models of the evolution of a photo-evaporating clump and the
shadowed region behind it were developed by~\citet{BerMcK90}
and~\citet{LefLaz94}, although these were mostly concerned with the
dense clump itself.  \citet{WilWarWhi01} used computational models to
study formation scenarios of the M16 pillars, setting up axi-symmetric
simulations with various initial density fields exposed to planar
ionising radiation.  They found that with their modelling assumptions,
multiple initial scenarios were capable of producing dense structures
resembling the ETs seen in M16.  These were found to be long-lived
quasi-equilibrium structures, raising the possibility that the pillars
could be quite old ($\ga 300\,$kyr).  Their work highlights the
difficulty in interpreting the observations -- a number of different
initial conditions and physical processes could produce pillars.  More
recently \citet{MiaWhiNelEA06} used an SPH code with radiative
transfer to model the photo-ionisation of a dense cloud, their results
supported the idea that the head of the pillar is in the early stages
of radiation-driven implosion.  Their modelling was of the head more
than the trunk, however, and it remains unclear whether a pillar could
form behind the imploding head on a time-scale as short as $100\,$kyr.
\citet{PouKanRyuEA07} modelled the photo-ionisation of a single dense
clump ($M=30\,\msun$) by an O star, specifically looking at the
shadowed region.  They found the shadow can produce a long neutral
tail of the dimensions of the M16 pillars, but without enough material
in the tail.  They note that adding in multiple clumps of different
sizes should increase the amount of tail material, as was found
by~\citet{LimMel03}, a suggestion we explore in detail in this paper.

The first global simulation of the expansion of an \hii\ region into a
turbulent density field was presented by \citet{MelArtHenEA06}. They
found that features such as ETs developed quite naturally due to the
uneven I-front expansion velocity, but due to the nature of the
simulation individual pillars are poorly resolved.
\citet{MacTorOisEA07} also modelled global expansion of an \hii\
region, but their simulations were more of the early I-front expansion
than of the later dynamical evolution. Very recently,
\citet{GriNaaWalEA09} modelled part of an expanding \hii\ region with
planar radiation impinging on a turbulent density field.  They also
found that pillar-like features developed naturally in their
simulations after about $250\,$kyr, but again the resolution in
individual pillars is poor.  \citet*{LorRagEsq09} also follow a
similar approach to form pillar-like structures.  They study the
angular momentum of dense clumps which form in their simulation and
find preferential alignment perpendicular to the direction of the
radiation field.  In work that is in some ways similar to ours, but on
much smaller scales, \citet{RagHenVasEA09} showed how
photo-evaporation flows from a large reservoir of dense gas can flow
into shadowed regions, recombine, cool, and begin to build up dense
pillar-like structures.

With some exceptions~\citep{WilWarWhi01,PouKanRyuEA07,RagHenVasEA09},
these works were not primarily focussed on how ETs form, and as a
result it is difficult to tell what physical processes are most
relevant.  The aim of this work is to focus on the shadowing mechanism
to see how effectively it can produce ETs in an idealised environment.
We have developed a radiation-magnetohydrodynamics code with which to
study this problem.  We will describe our code and algorithms in
section~\ref{sec:code}.  In section~\ref{sec:randomclumps} we describe
the initial conditions and show results from 3D simulations of the
photo-ionisation of a density field with randomly distributed dense
clumps.  These models show a number of structures resembling ETs which
form dynamically due to shadowing in the inhomogeneous medium.  In
section~\ref{sec:isolatedclumps} we simulate certain clump
configurations in isolation to demonstrate two different ways pillars
could form.  The first model has the clumps oriented almost like a
pillar in the initial conditions, whereas in the second model clumps
are swept past each other into a pillar-like structure.  We find that
neutral gas cooling has a strong effect on our results and in
section~\ref{sec:cooling} we repeat these two models using an
alternate thermal model with moderately strong neutral gas cooling.
The details and evolutionary time-scales change considerably, but the
formation mechanisms remain unchanged.  In
section~\ref{sec:discussion} we discuss the context and significance
of our results and in section~\ref{sec:conclusions} we present our
conclusions.

\section{Numerical Methods and Algorithms}
\label{sec:code}
\subsection{Fluid Dynamics}
We have written a modular, finite volume, fluid dynamics code to run
these simulations.  The code uses a uniform grid in 1, 2, or 3 spatial
dimensions with cubic cells. Scalar and vector fields are both
cell-centred.  The integrator for the fluid equations is based on the
algorithm described by~\citet*{FalKomJoa98}, which is dimensionally
unsplit and second order accurate in time and space.  We have separate
Riemann solvers for the Euler and Ideal Magnetohydrodynamics (MHD)
equations. We also add some artificial viscosity in a similar manner
to~\citet{FalKomJoa98} using a coefficient of $\eta_v=0.05-0.1$.
%
  This is required to ensure shocks travel at the correct speed
  in all directions.  It fixes for example the ``carbuncle problem''
  in the Double Mach Reflection test (\citealt{WooCol84} used the
  Lapidus viscosity prescription for this) and mitigates the Quirk
  instability~\citep{Qui94} for stationary grid-aligned shocks.
%

\subsection{Ray-tracing and Microphysics}
Our ray-tracing and microphysics routines are based largely on the
methods in~\citet{LimMel03} and on the \textit{C$^2$-ray} method
developed by~\citet{MelIliAlvEA06}. We use operator splitting to first
update the dynamics by a full timestep, and then run the microphysics
update over the full timestep. In this work we only consider
explicitly the ionisation of atomic hydrogen. We first describe the
ray-tracing algorithm and then the microphysics calculation.

The Short Characteristics tracer~\citep[e.g.][]{RagMelArtEA99} is used
to trace out rays from a source in a causal manner, calculating the
optical depth to a cell by interpolating between (previously
calculated) optical depths to neighbouring cells closer to the source.
Given that we are ignoring diffuse radiation (the On-the-Spot
approximation) the diffusion in the ray-tracer is not significant, and
is minimized using the weighting scheme given
by~\citet{MelIliAlvEA06}.

When the photo-ionisation time is short compared to other time-scales
(cooling, recombination, and collisional ionisation times) the
microphysics equations become difficult to solve explicitly so we
adopt a dual approach.  In cases of weak photo-ionisation, we use an
explicit 5th order Runge-Kutta technique with adaptive step-size to a
given relative accuracy~\citep{NR92}.  For strong photo-ionisation we
integrate explicitly until the Hydrogen ion fraction, $x$, satisfies
$x\geq0.95$, and then analytically integrate the equations assuming a
constant electron density (as described in~\citealt{MelIliAlvEA06}),
with bisection substepping to convergence (typically $2-4$ substeps).
For both of these methods we use a relative error tolerance of 0.001.

This algorithm also calculates the time-averaged optical depth through
the cell $\Delta\tau$, which is then used by subsequent cells in the
ray-tracer. \citet{MelIliAlvEA06} use a simple time average of
$\Delta\tau$, however we use a time average of $\exp(-\Delta\tau)$
since this gives a time average of the fraction of photons passing
through the cell. This can be easily seen in the (extreme) case of an
optically thick cell which is photo-ionised ``rapidly'' half way
through a unit timestep, so that
\begin{equation}
 \Delta\tau(t) = \left\{
  \begin{array}{cc}100&t<0.5\\0&0.5<t<1 \end{array} \right\} \,.
\end{equation}
The mean optical depth over the timestep is 50, but clearly half of
the incident photons will pass through the cell, and
$\int_0^1\exp(-\Delta\tau)dt=0.5$ gives the desired result. We do this
integration at the same time as the microphysics variables, to the
same accuracy criterion.

We use monochromatic radiation with a hydrogen photo-ionisation
cross-section of $6.3\times10^{-18}\,\mathrm{cm}^{2}$ and an energy gain
of $5.0\,$eV per photo-ionisation.  Collisional ionisation rates are
calculated with fitting functions from~\citet{Vor97}, and radiative
recombination (Case B) rates using the tables calculated
by~\citet{Hum94}.  The difference between planar radiation and
radiation from a point source can be quite significant if the size of
the computational domain is comparable to the distance to the source.
The rocket effect is weaker further from a point source due to the
inverse square law, which may extend the lifetime of any structures
that form.  This effect can, however, reduce the length of such
structures since the intensity of the radiation is higher at their
heads. In the case of M16, the heads of the pillars are about $2\,$pc
from the brightest star, and they are about $1\,$pc long, so the flux
dilution is more than a factor of 2 along their length. We therefore
use a point source in this work.

\subsection{Gas Cooling}
\label{subsec:cooling}
We use two cooling models in this work, denoted C1 and C2, which
differ in that C2 has significant cooling in the neutral gas.  Our
first model, \textbf{C1}, contains four elements:
\begin{enumerate}
\item Radiative losses due to recombining Hydrogen, calculated from
  the non-equilibrium ion fraction and temperature in each cell
  according to rates tabulated in~\citet{Hum94}.
\item Collisional ionisation of Hydrogen: this is relatively
  unimportant because the rates are typically very low, but we
  subtract the ionisation energy from the gas for each collisional
  ionisation.
\item Cooling due to heavy elements at high temperatures, using the
  collisional ionisation equilibrium (CIE) cooling curve tabulated
  in~\citet{SutDop93} and shown in their fig.~18.  This provides
  strong cooling in ionised gas with temperatures significantly larger
  than $10\,000\,$K.  
%
    Note that in CIE at $10\,000\,$K, Hydrogen, Nitrogen
    and Oxygen are neutral so this fitting function does not double
    count the other terms in our cooling function, at least for the gas
    temperatures encountered in our simulations.
%
\item A linear fit to collisionally excited emission from
  photo-ionised Oxygen and Nitrogen~\citep{Ost89}.
\end{enumerate}
The last term is the most important for this work, since these ionic
species are the dominant coolants in \hii\ regions and set the
equilibrium temperature in ionised gas of $T_{eq}\simeq8000\,$K. 
In experiments with different cooling functions for ionised gas, we found
that the most important factor for the dynamical evolution of our
models was the equilibrium temperature.  If the normalisation of the
cooling function is kept fixed at $8000\,$K, its slope has little
effect on the resulting dynamics so long as the slope is positive.  If
we had strong shocks in the ionised gas this aspect of the cooling
function would have more influence, but the photo-ionised gas in our
simulations has a very narrow temperature range.
 
In this prescription the neutral atomic gas has no efficient cooling
avenue, and shocked neutral gas is typically at $100-10\,000\,$K.
This is undoubtedly a limitation in our modelling, but we do not yet
model the formation of molecules, or the formation/destruction of
dust, which are the primary neutral gas coolants in star forming
regions.  To assess the effects of significant neutral gas cooling we
also use an alternate cooling function, \textbf{C2}, consisting of the
previous components in C1 plus additional exponential cooling
(Newton's Law) in neutral gas with a rate given by
\begin{equation}
  \dot{T} = \left[\frac{(1-x)^2}{10^N\mathrm{yrs}}\right]
  \left(T_{\infty}-T\right) \;,
\end{equation}
where $T$ is gas temperature, $T_{\infty}$ is the temperature to which
this cooling law relaxes at late times, $x$ is the ionisation fraction
of the gas, and $N$ is a parameter specifying the chosen cooling
time-scale, $t_c=10^N\mathrm{yrs}/(1-x)^2$.  The scaling with
$(1-x)^2$ ensures only mostly neutral gas is affected.  We set
$T_{\infty}=100\,$K and $N=4$ for the alternate models run in this
paper.  This is not an extreme model either in terms of the
equilibrium temperature or the cooling time, having less cooling in
dense gas than the model presented in~\citet{HenArtDeCEA09}.  It is a
very simple prescription, with an effect which is intermediate between
C1 and a two-temperature isothermal
model~\citep[e.g.][]{WilWarWhi01,GriNaaWalEA09,LorRagEsq09}.

\subsection{Code Tests}
\begin{figure*}
  \centering 
  \includegraphics[width=0.48\textwidth]{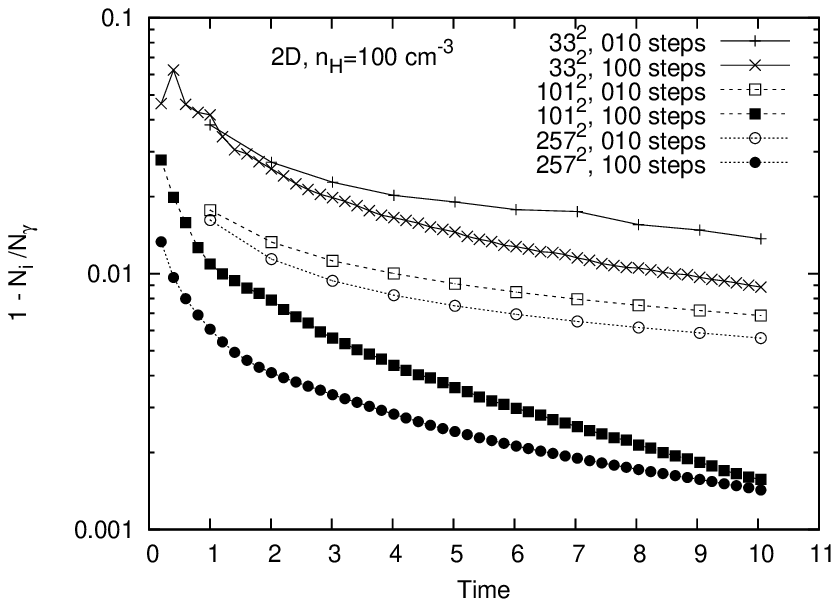}
  \includegraphics[width=0.48\textwidth]{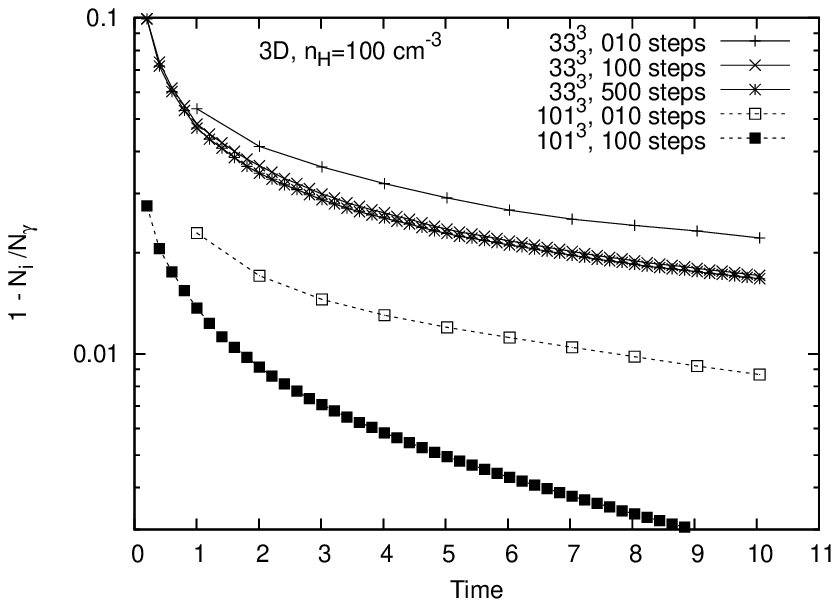}
  \includegraphics[width=0.48\textwidth]{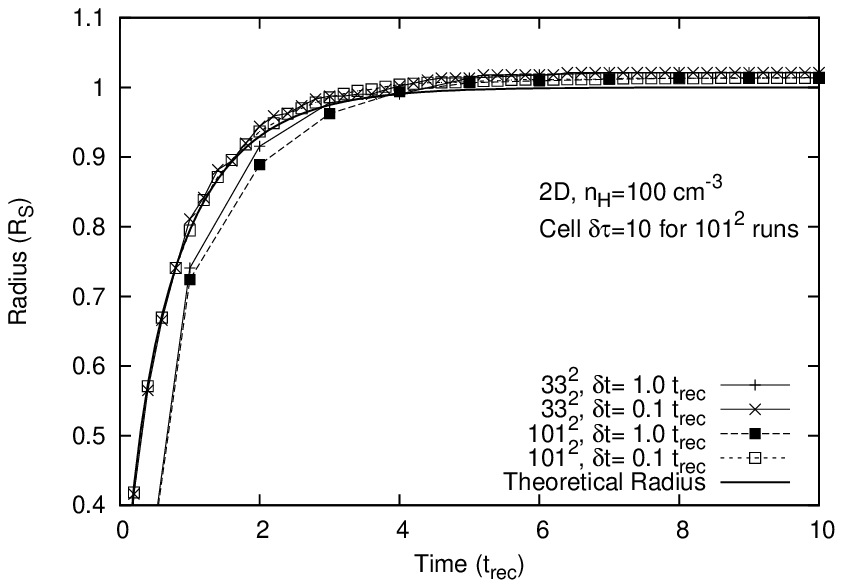}
  \includegraphics[width=0.48\textwidth]{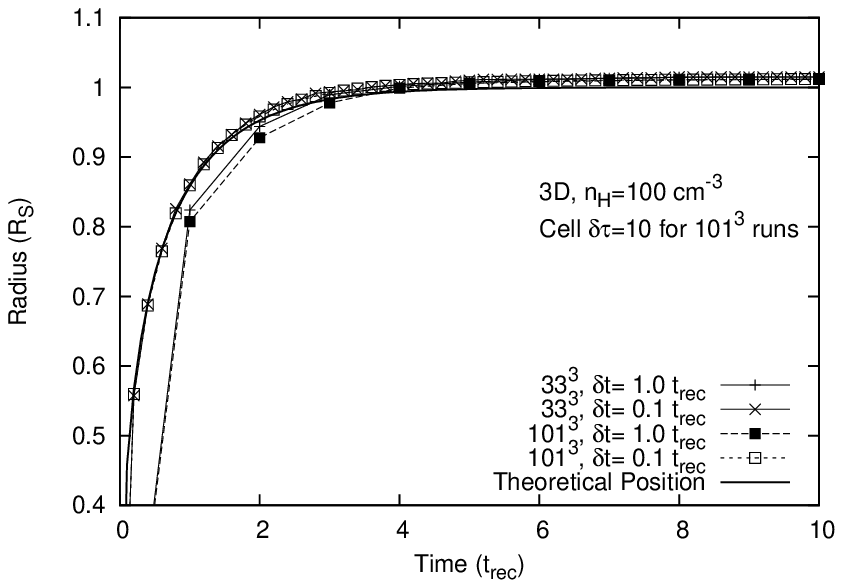}
  \caption{Tests of the radiative transfer algorithm in 2D (left) and
    3D (right).  The top two plots compare the number of ions to the
    number of photons emitted for a source in a uniform medium with
    dynamics and recombinations switched off.  The source is at the
    centre of the domain, and the I-front remains on the domain for
    the duration of the simulations.  This is a scale-free problem
    which we have run with parameters such that the cell optical depth
    is $\delta\tau=10$ for the runs with $101^2$ and $101^3$ cells
    ($\delta\tau=30.6$ for lower resolution and $3.93$ for higher
    resolution).  We always lose some photons due to the
    interpolation, but this decreases dramatically with resolution.
    The lower plots show the evolution of the I-front radius over
    time for runs with recombinations switched on.  Again it is a
    uniform medium with the same optical depths per cell.  The time is
    shown in units of the recombination time $t_{\mathrm{rec}}$, and
    the radius in units of the Str\"{o}mgren radius (or its 2D
    analogue).  When the timestep $\delta t=t_{\mathrm{rec}}$ the
    I-front propagates too slowly, but for $\delta
    t=0.1t_{\mathrm{rec}}$ it has the correct speed.}
  \label{fig:photon_cons}
\end{figure*}

We have extensively tested the fluid dynamics, microphysics, and
raytracing components of the code.\footnote{A brief description of our
  code and results from test problems can be found at
  \texttt{http://homepages.dias.ie/\~{}jmackey/jmac/}} For
hydrodynamics we used a range of shock-tube tests in
1D~\citep{Toro99}, and then in 2D at various angles to the grid axes,
with the code reproducing the correct solutions.  We have run the
double mach reflection test~\citep{WooCol84} and obtained good
agreement with the original work and with e.g.\ the \textsc{athena}
code~\citep[][fig.~16]{StoGarTeuEA08}.  Since we explicitly add in
numerical viscosity, the diffusion is slightly stronger than in the
\textsc{athena} code.  We have also done tests of implosions and blast waves in
2D and 3D, finding results consistent with previous
work~\citep[e.g.][]{StoGarTeuEA08}, and recovering the Sedov-Taylor
solution for the adiabatic blast wave in 3D.  We have also tested the
development of the Kelvin-Helmholz
instability~\citep[e.g.][]{AgeMooStaEA07}, finding very satisfactory
agreement with other work.

\citet{Ray79} calculated zero-dimensional shock models by following a
parcel of gas through a shock front.  We have tested our code against
his `Model E', a $100\kms$ steady shock with a $1\mu$G transverse
magnetic field, setting it up in our code as a 1D problem with 100\kms
gas hitting a dense cold layer and allowing the system to relax to an
equilibrium state. Our test reproduced well the ion-fractions for
Hydrogen and Helium, the gas temperature and the density as a function
of position.  We have also run models at higher shock velocity and
find that for $v\ga130\kms$ the shocks become overstable, in agreement
with previous work~\citep*[e.g.][]{InnGidFal87}.

Photo-ionisation was tested in conjunction with raytracing using
similar tests to those in~\citet{MelIliAlvEA06}, where the dynamics is
switched off.  We started with 1D rays from a source at infinity,
without dynamics or recombinations.  For a grid with 1\,000 cells, we
computed models with cell optical depths
$\Delta\tau=\{0.1,1,10,100\}$, and where the total number of timesteps
varied from $t_{\mathrm{sim}}/\delta t = \{10^1,10^2,10^3,10^4\}$.  The
error in I-front position compared to the analytic value was found to
converge rapidly to less than one cell width with increasing time
resolution.  For models with recombinations turned on, errors were no
more than than one cell width for all runs with $>10$ timesteps per
recombination time, except for low density models where the I-front is
resolved.

In 2D and 3D, we computed the expansion of circular and spherical
I-fronts from a point source into a static medium, with and without
recombinations.  Without recombinations, the models provide a test of
photon conservation (by comparing the number of ions to photons
emitted as a function of time).  We plot the photon conservation in
the top two panels of Fig.~\ref{fig:photon_cons}, for 2D on the left
and 3D on the right.  These figures show the relative sizes of
ray-tracing and time-integration errors as a function of resolution
and dimensionality.  For the very low resolution runs ($33^2$ and
$33^3$ cells) we lose between 1 and 10 per cent of photons due to
interpolation errors in the ray-tracing when the ionised region is
$<10$ cells across.  With increased spatial resolution the errors
decrease strongly whereas increased time resolution doesn't help the
33 cell runs significantly.  There is a dramatic improvement in
accuracy with time resolution for the 101 and 257 cell runs.  These
results show that the errors are interpolation dominated when the
number of cells is much smaller than the number of timesteps and
time-integration dominated in the opposite limit.  Using the weighting
scheme recommended by~\citet{MelIliAlvEA06} we find that I-fronts are
circular to within a cell width over a wide range of densities,
luminosities, and spatial and temporal resolutions.

The lower panels of Fig.~\ref{fig:photon_cons} show the position of
the I-front as a function of time for simulations with recombinations
included, modelling the idealised Str\"{o}mgren Sphere analysis and
its 2D analogue.  The mean I-front position is always within $1-2$ per
cent of the analytic value except at very early times when it has only
crossed a few cells, or when the timesteps are of order the
recombination time, $t_{\mathrm{rec}}=(\alpha n_{\mathrm{H}})^{-1}$,
where $\alpha$ is the (case B) recombination coefficient set to a
constant for this test
($\alpha=2.59\times10^{-13}\,\mathrm{cm}^{3}\,\mathrm{s}^{-1}$), and
$n_{\mathrm{H}}$ is the Hydrogen number density.  This is an expected
limitation of the C$^2$-ray method since it uses time-averages of the
photon flux through each cell~\citep[see][]{MelIliAlvEA06}.  For the
tests where $\delta t=t_{\mathrm{rec}}$ we underestimate the I-front
velocity while it expands to the Str\"{o}mgren radius.  The error is
slightly larger at higher spatial resolution because we have to do the
same inaccurate time-average across more cells and the error is always
on the side of losing photons.  For sufficient time resolution,
however, the I-front propagates at the correct speed, and it is worth
noting that the photo-ionisation time for a cell is much shorter than
the recombination time while the I-front is expanding rapidly.  We do
not need to resolve this timescale to get accurate results.  This is
the major strength of the C$^2$-ray algorithm.

We have also used our code to reproduce the simulations
of~\citet{LimMel03}, where dense clumps in different configurations
were photo-evaporated by planar ionising radiation.  We find our code
gives largely consistent results despite significant differences in
the numerical methods, e.g.~the~\citet{LimMel03} simulations did not
use an explicitly photon-conserving algorithm.

\section{Random Clumps Simulations and Results}
\label{sec:randomclumps}

\begin{table*}
  \begin{tabular}{  | r l l l l l l l l || l |}
    \hline
    No. & $n_{cl}$ & $M_{cl}$ & $N_{cl}$ & Size  & $M_{i}$(min/max) &
    $n_{\mathrm{max}}$ & $M_J$ & Flux & Results \\
    \hline \hline
     1 & $10^3$ & 72.2 & 115 & C & $0.11-1.1$ & $2.7\times10^5$ & 12.4 & $10^{10}$ & Nothing like ETs; corrugated I-front. \\ 
     2 & $10^3$ & 72.2 & 115 & C & $0.11-1.1$ & $2.7\times10^5$ & 12.4 & $10^{11}$ & Nothing like ETs; corrugated I-front. \\ 
     3 & $10^3$ & 72.2 & 103 & E & $0.36-1.1$ & $2.0\times10^4$ & 45.3 & $10^{11}$ & Too diffuse to form dense structures. \\ 
     4 & $10^3$ & 72.2 &  40 & C & $0.36-2.9$ & $9.1\times10^5$ & 6.7  & $10^{10}$ & Low mass and low density ETs form.  \\ 
     5 & $10^3$ & 72.2 &  40 & C & $0.36-2.9$ & $9.1\times10^5$ & 6.7  & $10^{11}$ & Short lived Cometary structures. \\ 
     6 & $10^3$ & 72.2 &  40 & E & $1.3-2.6$  & $4.1\times10^4$ & 31.7 & $10^{11}$ & No dense ET-like structures. \\ 
     7 & $10^3$ & 72.2 &  54 & C & $0.22-2.2$ & $4.8\times10^5$ & 9.3  & $10^{10}$ & Nothing dense enough or trunk-like. \\ 
     8 & $10^3$ & 72.2 &  54 & C & $0.22-2.2$ & $4.8\times10^5$ & 9.3  & $10^{11}$ & Nothing dense enough or trunk-like. \\ 
     9 & $10^3$ & 72.2 &  85 & C & $0.14-1.4$ & $4.0\times10^5$ & 10.1 & $10^{10}$ & Small ET-like structures. \\ 
    10 & $10^3$ & 72.2 &  85 & C & $0.14-1.4$ & $4.0\times10^5$ & 10.1 & $10^{11}$ & Small cometary structures, low density. \\ 
    11 & $10^4$ & 666 & 118 & C & $1.0-10$ & $3.5\times10^6$ & 3.4   & $10^{11}$ & Some ETs among large `mountains'. \\ 
    12 & $10^4$ & 666 &  45 & C & $3.3-27$ & $7.6\times10^6$ & 2.3   & $10^{11}$ & Multiple dense long-lived ETs. \\ 
    13 & $10^4$ & 666 &  40 & E & $12-24$ & $5.7\times10^5$ & 8.5  &   $10^{11}$ & Large `mountains' with a few dense ETs. \\ 
    14 & $10^4$ & 666 &  59 & C & $2.0-20$ & $5.1\times10^6$ & 2.8   & $10^{11}$ & Many dense ETs. \\ 
    15 & $10^4$ & 666 &  83 & C & $1.3-13$ & $3.9\times10^6$ & 3.3   & $10^{11}$ & Many dense ETs. \\ 
    16 & $4\times10^3$ & 266 & 40 & E & $2.4-10.4$ & $1.8\times10^5$ & 15.1 & $10^{11}$ & Formed long-lived ETs, see Fig.~\ref{fig:RC3D_volnH}. \\
    17 & $n_b=200$ & 84.9 & 3 & $0.1\,$pc & 28.3 & $1.0\times10^5$ & 20.1 & $2.9\times 10^{11}$ & Almost co-linear clumps, see Fig.~\ref{fig:M101std_Vol}. \\
    18 & $n_b=200$ & 83.7 & 3 & $0.1\,$pc & $27.7-28.3$ & $1.0\times10^5$ & 20.1 & $2.9\times 10^{11}$ & Triangular configuration, see Fig.~\ref{fig:M112std_Vol}. \\
     \hline
  \end{tabular}
  \caption{Simulation parameters for the test 3D simulations ($1-15$)
    and the one high resolution model (16).   All models have a
    background density $n_b=100\,\pcmc$ and triaxial clumps as
    defined in the text, except for models 17 and 18 which have
    $n_b=200\,\pcmc$ and spherical clumps (see
    section~\ref{sec:isolatedclumps}).  Clump density, $n_{cl}$ refers
    to the mean density of gas put into clumps ($\pcmc$) and the next
    column $M_{cl}$ gives the total mass in clumps (all masses are in
    solar masses, $\msun$).   
    $N_{cl}$ lists the number of clumps in each model.  For
    size, `C' refers to compact and `E' to extended, as described in
    the text (section~\ref{sec:2Dsims}).  $M_{i}$(min/max) gives the
    range of clump masses for the model.
    The Jeans mass (defined as $M_J=203\msun\,T^{1.5}n^{-0.5}$ for
    adiabatic gas) is calculated for the densest point in the initial
    conditions ($n_{\mathrm{max}}$) and for a temperature of $10\,$K.
    We do not use the actual clump temperatures because the initial
    constant pressure state has very low temperatures at the densest
    points ($T\sim1\,$K in some cases); this gives a meaningless Jeans
    mass since if we included gravity we wouldn't need to impose such
    an artificially low temperature.
    The Flux is the monochromatic ionising photon flux
    ($\pcmsps$) entering the domain.
    All models use clumps with random positions and properties (within
    ranges) except models 17 and 18 which are set up to model specific
    clump configurations in isolation.
  }
  \label{tab:3Dsims}
\end{table*}

Previous
studies~\citep[e.g.][]{MelArtHenEA06,MacTorOisEA07,GriNaaWalEA09} have
used a turbulence model to generate a density field into which the
ionising radiation propagates.  In this work we wish to isolate the
shadowing effect of the radiation from other rather uncertain physics
such as the type of turbulent motions we inject into the gas.  In this
respect our work is more similar to that of~\citet{WilWarWhi01}.  We
add dense clumps to a uniform density field in the following way:
\begin{enumerate}
\item We start by setting the mean background density, $n_b$, ranging
  from $10$ to $10^3\pcmc$ in different simulations.
\item We choose a total mass to put into clumps by calculating the
  mass associated with a smoothed mean density in clumpy material,
  $n_{cl}$, in a subset of the full simulation domain.  The subset is
  chosen to keep clumps away from simulation boundaries, particularly
  the boundary nearest the radiation source.
\item We set the maximum and minimum mass and size of the clumps and
  draw clumps randomly within these limits until all the mass in
  clumpy material has been used up. The size of a clump in each
  direction is chosen separately, so typically we get triaxial clumps.
\item The clump's mass and size determine its peak overdensity for a
  given radial profile.  In this work we use Gaussian profiles.
\item The clump's position is then set at random on the domain, within
  the allowed region, and it is rotated by a random angle in all three
  directions.  Clumps are added to the background density field one by
  one, in a linear superposition when they overlap.
\item The initial conditions are static everywhere, and we assign a
  constant pressure throughout.
\end{enumerate}

The method of random sampling can strongly influence the distribution
obtained.  Clump positions, radii, orientations, and masses are
selected randomly within certain minimum and maximum values.  In all
cases, we select a random number on a linear scale between the two
limits.

\subsection{2D Simulations}
\label{sec:2Dsims}
The above procedure can also be performed in 2D with slab symmetry,
enabling us to do a study with a wider range of parameters than is
possible in 3D.  We performed 184 2D simulations on a
$1.5\times1.5\,$pc$^2$ domain with a radiation source $2\,$pc off the
domain in one direction.  The background densities used were
$n_b=\{10,100,1000\}\,\pcmc$; the ionising photon fluxes entering the
domain were $F_{\gamma}=\{10^{10},10^{11},10^{12},10^{13}\}\,\pcmsps$;
the number of clumps were $N_{cl}=\{10,30,100\}$, being either compact
or extended, round or triaxial, with scale radii $0.0225\,\mathrm{pc}
\leq R_s \leq 0.15\,\mathrm{pc}$.  Compact round clumps had radius
$0.05\,$pc, extended round clumps were $0.12\,$pc, compact triaxial
clumps were $[0.0225-0.06]\,$pc, and extended triaxial clumps were
$[0.06-0.15]\,$pc.  The total mass in clumps was chosen by the above
method, setting a smoothed mean density in clumps of
$n_{cl}=\{10^2,10^3,10^4\}\,\pcmc$ and calculating the associated
mass.  For the 2D models we set the maximum and minimum clump masses
to be very close to each other so that all clumps would have similar
masses.

In addition to the 184 models with the parameters mentioned above, we
also selected particular models and varied the random seeds, also
using larger domains to ensure boundary effects were not important.
For models with 10 clumps, the specific clump configuration was
important, but for 30 and 100 clumps, the random seed had little
effect on whether or not the model produced ETs.  The boundary effects
were found to be minimal.  We use outflow, or zero-gradient, boundary
conditions for all of the simulations in this paper.  This is the
logical choice for the lateral boundaries and for the boundary
furthest from the source.  For the boundary nearest the source a
reflection boundary could be imposed to model a pressure confined
\hii\ region, but we have chosen to model an \hii\ region which has
broken out of its parent molecular cloud and is no longer pressure
confined in at least some directions.

Care must be taken with the transition from R-type to D-type I-fronts.
If this occurs too close to the boundary nearest the source spurious
inflows can be set up by rarefaction waves propagating backwards from
the stalling I-front.  This can introduce an unphysical aspect to the
simulation results, therefore we chose all of our simulation
parameters such that the I-front in the background medium remained
R-type well into the domain.  This ensured that photo-evaporation
flows dictated the flow at the boundary rather than vice versa.  We
have verified that the R-type to D-type transition occurs as expected
with a shock driven ahead of the I-front into the neutral gas and a
strong photo-evaporation flow established in the opposite direction.

Of the 184 2D models, approximately 30 produced pillar-like
structures, with another $\simeq20$ producing short features more like
heaps than pillars, and another $\simeq10$ producing very short-lived
pillar-like features lasting $<50\,$kyr.  The most successful groups
of models had background and clump densities of
$[n_b,n_{cl}]=[10^2,10^3]\pcmc$, $[10^2,10^4]\pcmc$, and
$[10^3,10^4]\pcmc$, with 10 or 30 clumps.  Those models which
successfully produced pillars had a combination of large low-density
regions where the I-front could propagate far from the source, and
large dense clumps which were capable of stalling the I-front for
sufficient time to produce a pillar in the shadowed region.  The
trunks of the pillars were formed from clump material more than
ambient gas, simply because the ambient gas is not dense enough.  If
the background density was high enough to contribute significantly to
a pillar, then it was also high enough to stall the I-front and a
pillar could not form.  The clump material which made up a pillar
always came from several clumps; a lone clump in these models never
gathered enough ambient gas into the tail to produce a pillar, nor did
it contribute enough of its own gas to the shadowed region to form the
trunk of a pillar.  Typically, pillars formed via the interaction of a
number of clumps shadowing each other and accelerating past/through
each other produced a dense tail which lasted anything from
$50-200\,$kyr, depending on the inertia of the leading clump.

\subsection{3D Test Simulations}
There are significant differences between the 2D and 3D cases: the
stellar flux drops off more rapidly with distance in 3D; shocks in the
shadowed regions are converging or diverging whereas in 2D they are
planar; there is also an extra dimension for gas to flow past a clump.
These favour the formation of ET-like features in 2D, however our 2D
results can be used to constrain our choice of parameters for more
computationally expensive 3D simulations.  Some of these differences
can be modelled by comparing 2D slab-symmetric and axi-symmetric
models.  For our models, however, it is the asymmetries which generate
the ETs and these cannot be modelled in axi-symmetry.

\begin{figure*}
  \centering 
  \includegraphics[width=1.0\textwidth]{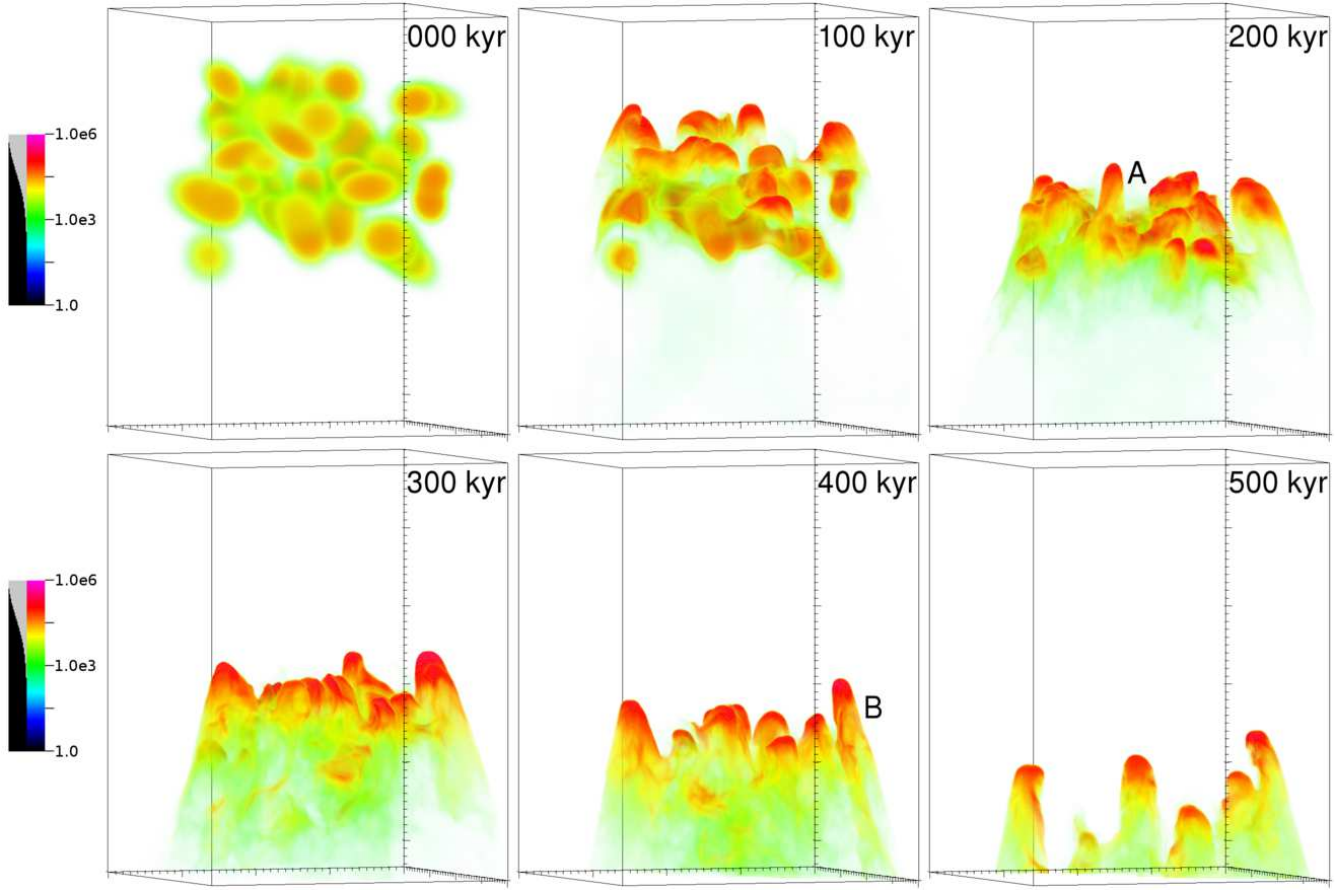}
  \caption{Volume Rendering of a 3D simulation of the photo-ionisation
    of a random clump distribution.  Neutral gas density is plotted on
    a log scale as indicated (in units of $\pcmc$), with gas at
    $n_{\mathrm{H}} \la 10^4\pcmc$ made transparent to show the
    difference between neutral tails and dense pillars.  Figures are
    shown for outputs every $100\,$kyr from 0 to $500\,$kyr.  The
    source is $2\,$pc off the top of the domain.  Note the pillar
    labelled `A' front and centre in the third panel, and another
    labelled `B' at right back in the later panels.}
  \label{fig:RC3D_volnH}
\end{figure*}

Initially we computed fifteen 3D models with $168\times128^2$ grid
cells, since the 2D models had shown that this was sufficient to
resolve the larger features in the simulations.  The physical domain
was $2.625\times2\times2\,\mathrm{pc}^3$.  The parameters we varied in
each model are listed in Table~\ref{tab:3Dsims}, models $1-15$.  All
models start with a constant initial pressure of
$p_g/k_B=2\times10^{5}\,\mathrm{cm}^{-3}\,\mathrm{K}$ where
$k_B=1.38\times10^{-16}\,\mathrm{erg}\,\mathrm{K}^{-1}$ is the
Boltzmann constant; this pressure is higher than the typical ISM value
and is a compromise between not having the temperature too high in
diffuse gas and too low in dense gas.  The results are not sensitive
to this initial pressure because the ionised gas pressure is
$100\times$ greater and it is the pressure difference which drives the
dynamical evolution of our models.  All 3D models had $n_b=10^2\pcmc$;
of these four had
$\{n_{cl},F_{\gamma}\}=\{10^3\pcmc,10^{10}\pcmsps\}$, six used
$\{10^3\pcmc,10^{11}\pcmsps\}$, and five used
$\{10^4\pcmc,10^{11}\pcmsps\}$.  Of these, the models with
$n_{cl}=10^3\pcmc$ had too little mass in clumps and none of the
models produced anything as dense or massive as a pillar.  The models
with $n_{cl}=10^4\pcmc$, by contrast, did produce massive and
long-lived pillars, simply because the initial clumps were so massive
and dense.  The models with fewer clumps were more successful.  With
too many clumps there were no large low-density regions, resulting in
structures more like mountains than long columns.  These dense models
were rather unrealistic however, with initial densities in clumps of
$n_{\mathrm{H}}\ga 10^6\pcmc$, very gravitationally unstable, as can
be seen by comparing clump masses to Jeans masses in
Table~\ref{tab:3Dsims}.

It was difficult to devise an automatic method for identifying
candidate ETs in the 3D simulations.  Instead we generated images of
the volume renderings from a number of viewing angles and picked out
structures with $n \geq 10^4\,\pcmc$, an aspect ratio $\gtrsim2$, and
which persisted for $>50\,$kyr.  Any candidates were then studied more
closely with cross-section and isosurface plots to verify that they
could be classified as ETs.

\subsection{Large 3D Simulation}
Based on these tests we tried an intermediate model (Model 16 in
Table~\ref{tab:3Dsims}) with less extreme initial densities.  We chose
$n_{cl}=4\times10^3\pcmc$, intermediate between two models in the
tests, with clumps restricted to the central 75 per cent of the domain
in $y$ and $z$ directions, and to 45 per cent of the longer $x$
domain.  This gave a mass in clumps of $266\,\msun$, making up
$\simeq90$ per cent of the total gas mass.  This was distributed among
40 clumps ranging in mass from $2.4-10.4\,\msun$, and in radius from
$0.06-0.14\,$pc.  The peak density in the initial conditions was
$n_{\mathrm{H}}\simeq 1.8\times10^5\pcmc$, and the background density
was $n_b=100\pcmc$.  The ionising source was placed $2\,$pc off the
domain in the long direction and had an ionising flux of
$10^{11}\pcmsps$ at the front edge of the grid.  This model
successfully produced three ET-like structures and we repeated it with
$512\times384^2$ grid cells with a similar physical size of
$2.67\times2\times2\,\mathrm{pc}^3$.  A volume rendering of the neutral
gas number density is shown in Fig.~\ref{fig:RC3D_volnH} at $100\,$kyr
intervals from the initial state to $500\,$kyr.  The number density is
on a log scale, with gas at $n_{\mathrm{H}} \la 10^4\pcmc$ made
transparent (the transfer function is shown beside the log scale).
This highlights only the densest structures to distinguish between
shadowed regions and dense tails.

After $100\,$kyr the front clumps have begun to accelerate away from
the source and have clearly been compressed significantly.  The first
pillar (labelled `A') forms near the front of the domain after about
$170\,$kyr and lasts for about 50-70kyr before being accelerated back
to the main body of neutral gas.  It is about $0.5\,$pc in length with
an aspect ratio of $2-3$.  This frame also shows a large mass of
merged clumps at the far right.  By the next frame at $300\,$kyr, this
mass has begun to look more like a pillar, along with another mass at
the far left.  At $400\,$kyr both of these structures clearly resemble
ETs, with the one on the right (labelled `B') slightly more developed
and measuring about $0.75\,$pc in length.  All neutral gas coloured
yellow to red has a number density $n_{\mathrm{H}}\ga10^4\pcmc$, so
these are dense enough to be considered pillars and not just shadows.
The pillars are formed almost entirely from dense clump material that
has been pushed into the shadowed regions.  In the last frame at
$500\,$kyr pillar B has a surprisingly similar morphology to the
largest pillar of the three in M16, leaning over its neighbours.  This
pillar has survived for $200\,$kyr in the simulation, and is likely to
live for significantly longer.

\begin{figure*}
  \centering 
  \includegraphics[width=1.0\textwidth]{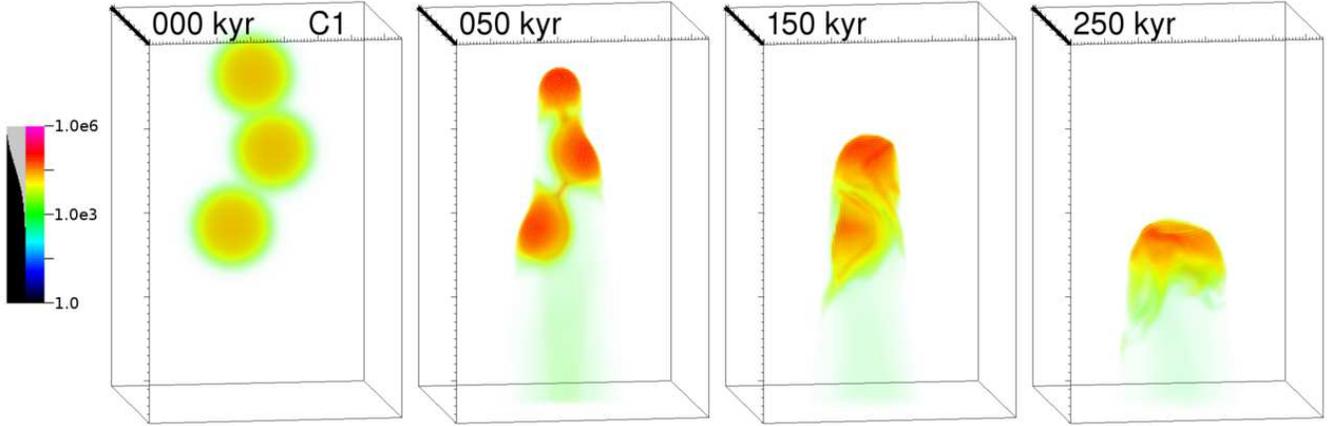}
  \caption{Volume Rendering of a 3D simulation of isolated pillar
    formation with three massive clumps initially almost in a line,
    using cooling model C1.  Neutral gas number density is plotted
    ($\pcmc$), with low density gas made transparent to show the
    difference between low density neutral tails and dense pillars.
    The transfer function is shown beside the log scale to the left.
    Simulation times are indicated on each panel.}
  \label{fig:M101std_Vol}
\end{figure*}

We tested the numerical convergence of these results by comparing the
structures that formed when this calculation was performed at
resolutions of $128^3$ and $256^3$ (with shorter $x$ domains) with the
simulation above.  Positions, sizes, and lifetimes of the pillars were
found to be consistent at all resolutions.  Transient peak densities
increased with resolution, as expected in problems involving strong
compression.

We summarise the basic results from these 3D models as follows.  It is
easier and more common for a configuration of clumps to produce a
structure like a mountain or corrugation than an ET.  In particular,
in our 3D models we required clumps of a few solar masses in a density
field that had substantial low-density regions surrounding the clumps
in order to generate ET-like structures.  Our best model produced a
single short-lived clump after about $170\,$kyr, and two or three
longer-lived pillars after $300\,$kyr.  No models produced an ET in
less than $150\,$kyr.  The pillars we formed were $0.5-0.75\,$pc long,
had number densities of $\ga10^5\pcmc$ at the densest part of the
head, and of $1-4\times10^4\pcmc$ in the denser parts of the trunks.
These properties are similar to observed ETs.  At $400\,$kyr, there
was $110\,\msun$ of neutral gas in the domain, of which $65\,\msun$
had $n_{\mathrm{H}}>10^4\,\pcmc$, divided among two ETs and the other
clumps.  This gives somewhat lower masses in our ETs than in the M16
pillars (masses $\sim30-90\,\msun$), but comparable to estimates for
other observed ETs~\citep[e.g.][]{GahCarJohEA06}.

\begin{figure*}
  \centering 
  \includegraphics[width=0.83\textwidth]{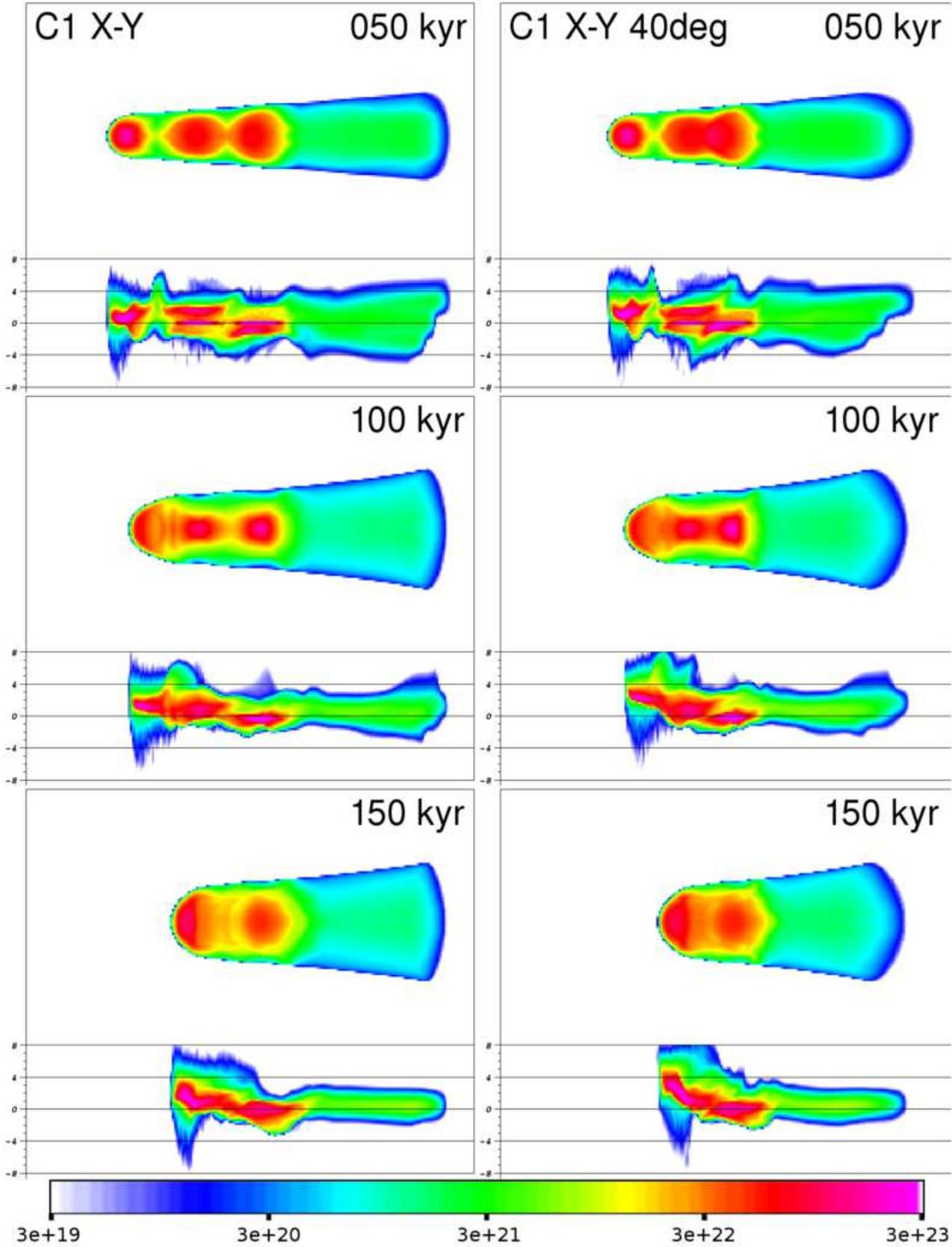}
  \caption{Projections through the ``co-linear clumps'' 3D simulation
    shown in Fig.~\ref{fig:M101std_Vol} with cooling model C1.  Gas
    column density of neutral (atomic) hydrogen is plotted in a 2D
    projection above and a position-velocity diagram below (with
    radial velocities indicated in units of~\kms).  Times are
    indicated on each panel.  The normal vector for the projections is
    at angles of 20\degr\ and 40\degr\ from the $z$-axis in the left
    and right sequences respectively, and is perpendicular to the
    $y$-axis.  Column density is shown on a log scale as indicated (in
    units of $\pcms$).}
  \label{fig:M101std_proj}
\end{figure*}

\section{Analysis of Clump Configurations}
\label{sec:isolatedclumps}
Certain configurations of clumps which occurred in the random initial
conditions gave rise to pillar-like structures.  To investigate how
these structures are formed, we have simulated configurations of
clumps in isolation to model the formation of pillars A and B.  These
are models 17 and 18 in Table~\ref{tab:3Dsims}.  The calculations used
a computational domain with $192\times128^2$ zones covering
$2.25\times1.5^2\, \mathrm{pc}^3$, a source at $-2\,$pc in the
$x$-direction with a luminosity in ionising photons of
$L_{\gamma}=2\times10^{50} \,\mathrm{s}^{-1}$ and an ambient gas density
of $200\pcmc$.  We have also run these simulations with 49 small
random clumps superimposed on this, corresponding to an extra mean
density of $400\pcmc$, but found very little difference in the
results.  Superimposed on this density field are three massive clumps
of up to $28.3\,\msun$ each, with a peak density of up to
$n_{\mathrm{H}}=10^5\pcmc$ (overdensity of 500) and Gaussian profiles
with scale radius $0.09\,$pc.  We have scaled up the mass and radius
of the clumps compared to the previous simulations in order to get
closer to the masses and sizes of the M16 pillars, which are
$\simeq30-90\,\msun$.  As expected for simulations without
self-gravity, the results with less massive clumps were very similar
except that the time-scales were shorter.  In the results that follow
we vary the relative positions of the three clumps. The ionising flux
at the front side of the clumps is about $2.9\times10^{11}\pcmsps$.
The gas is initially neutral, with constant pressure
$p_g/k_B=2\times10^{5}\,\mathrm{cm}^{-3}\,\mathrm{K}$, corresponding
to $1000\,$K in the lowest density gas.

\begin{figure*}
  \centering 
  \includegraphics[width=1.0\textwidth]{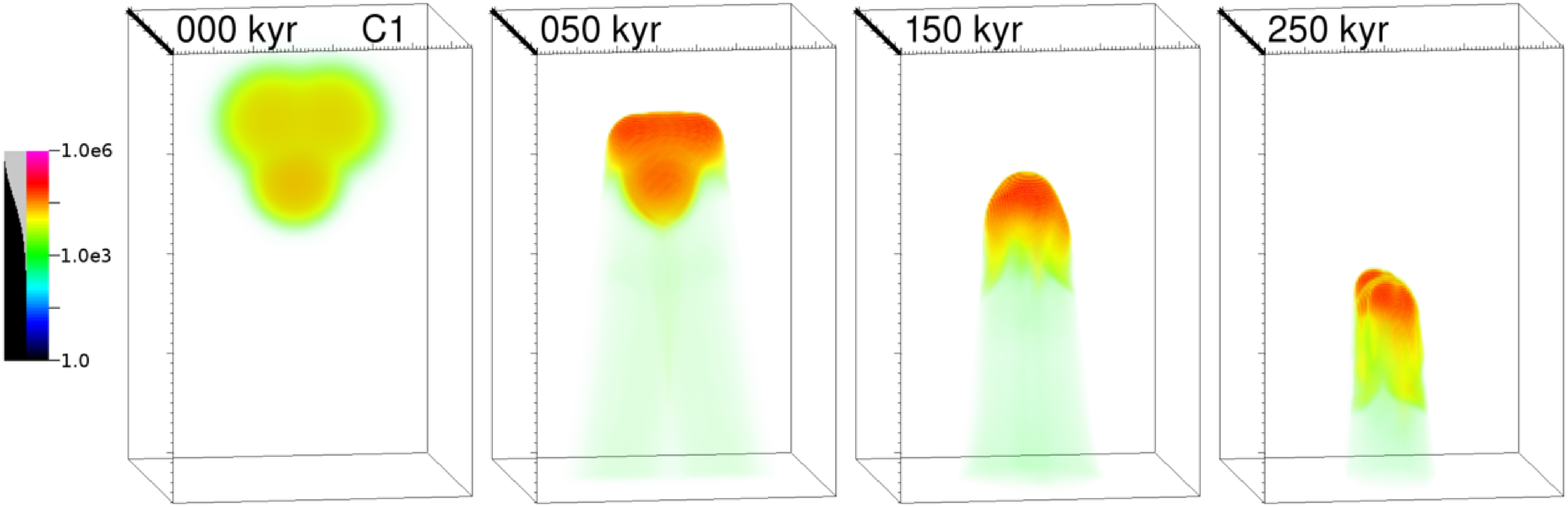}
  \caption{Volume Rendering of a 3D simulation of isolated pillar
    formation with three massive clumps initially in a triangle
    configuration, using cooling model C1. Neutral gas number density
    is plotted ($\pcmc$), with low density gas made transparent to
    show the difference between low density neutral tails and dense
    pillars.}
  \label{fig:M112std_Vol}
\end{figure*}

\subsection{Three almost co-linear clumps (Pillar B)}
We first investigate the effect of several clumps partially shadowing
each other in a roughly linear fashion (model 17), similar to the
configuration that formed pillar B. The three massive clumps are
located in the same y-plane as the source at positions
$[0.30,0.72,0.72]$, $[0.75, 0.72, 0.84]$, and $[1.2, 0.72, 0.60]$
(measured in parsecs from a corner of the domain nearest the source).
The front clump thus partially shadows the two behind it.

Volume renderings for the run using cooling model C1 are shown in
Fig.~\ref{fig:M101std_Vol} at times 0, 50, 150, and 250 kyr.  At later
times the clumps merge and the structure no longer resembles a pillar.
Projections at angles of 20\degr\ and 40\degr\ to the $z$-axis (where
0\degr\ is perpendicular to the pillar) are shown in
Fig.~\ref{fig:M101std_proj}, with column density shown in the maps,
and line of sight (LOS) velocity shown in position-velocity ($P-V$)
diagrams below the maps.  The 3D $P-V$ datacube is projected on to 2D
by summing the contribution of all the pixels in a given image
$y$-column at each $x$ position.  This means the normalisation of the
$P-V$ column densities is on a somewhat arbitrary scale.

Fig.~\ref{fig:M101std_Vol} shows how the partially shadowed clumps
compress and move sideways into the shadow, while the fully exposed
clump is accelerated and merges into them.  In this situation the gas
is initially aligned as a pillar-like structure, and the compression
due to photo-ionisation of the surroundings serves to enhance this
appearance for a limited time.  The neutral gas only resembles a
pillar for $150\,$kyr; this structure is a relic from the initial
conditions as opposed to being generated by dynamical evolution.  The
$P-V$ diagrams in Fig.~\ref{fig:M101std_proj} clearly show gas moving
at very different speeds away from the radiation source, so it is
likely that some of this gas will move into the tail as the structure
recedes from the source and is subjected to lower ionising fluxes.

Even though model 16 did not show such a clear
early-forming pillar, it is nevertheless interesting that the
projected column densities in Fig.~\ref{fig:M101std_proj} show
structures that resemble pillars for the first $150\,$kyr of the
simulation.  The clumps were not as neatly arranged in the random
clumps model, but given that the ISM in molecular clouds tends to be
filamentary and clumpy, it is certainly possible that there are some
ETs in \hii\ regions which are formed in this way i.e.\ purely from
initial conditions and not dynamically.  This is a less satisfactory
explanation for M16 however, since we would need three lines of
clumps/filaments positioned beside each other, and all pointing by
chance back to the brightest star in the nebula.

Looking at the evolution in more detail, the $P-V$ diagram shows the
acceleration of the first clump away from the source, up to about
$6\,\kms$ in $150\,$kyr (The true velocity is
$v_{\parallel}/\sin{\theta}$).  For all of this time the shadowed gas
remains essentially stationary.  The velocity signature of this
formation mechanism is very clear, with the head of the pillar
receding from the star faster than the trunk.  This is the opposite of
what is seen in other formation scenarios such as the model
of~\citet{LefLaz94} of a tail forming behind a single clump which has
the tail streaming away from the star faster than the head.  We show
the two perspectives at 20\degr\ and 40\degr\ to demonstrate that even
though 20\degr\ is close to perpendicular, they both show the same
trends in velocity along the pillar (In subsequent figures we only
show 20\degr\ projections).

We note, however, that the almost linear decrease in LOS velocity from
the head to the tail is significantly enhanced by transverse motions
in the gas.  The middle clump is being pushed away from the observer
into the shadow, and the right-most clump towards us (cf.\
Fig.~\ref{fig:M101std_Vol}).  If we make the same projections from the
opposite side of the simulation these transverse velocities are
reversed leading to a much less obvious velocity gradient (although
the large velocity at the head is still evident).  We return to this
issue later in Section~\ref{sec:cooling} where we show velocity
profiles from different perspectives.

\subsection{Triangle of three clumps (Pillar A)}
\begin{figure*}
  \centering
  \includegraphics[width=0.83\textwidth]{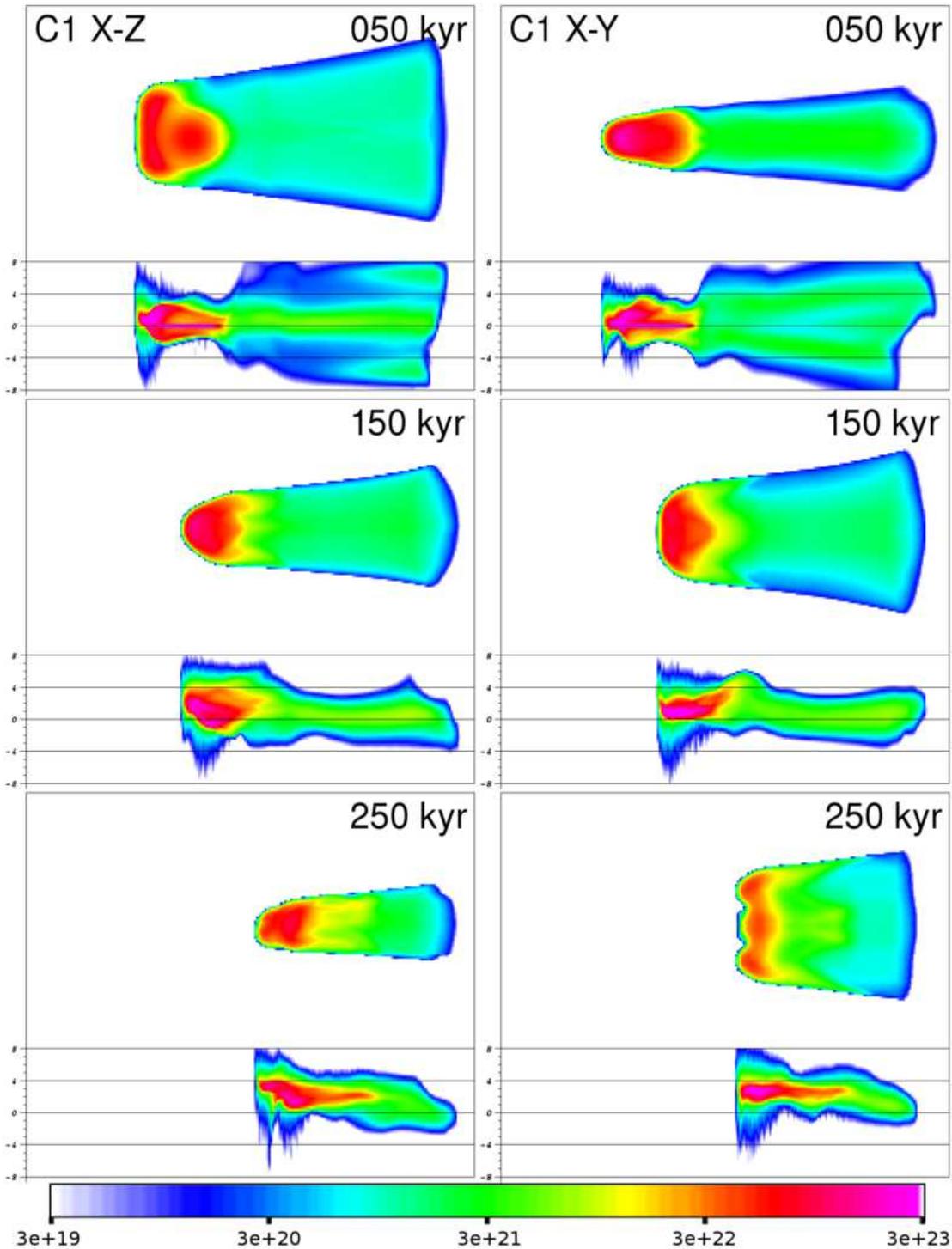}
  \caption{Projections through the ``triangle of clumps'' 3D
    simulation shown in Fig.~\ref{fig:M112std_Vol} with the C1 cooling
    model.  The projection is at 20\degr\ to the $y$-axis on the left,
    and to the $z$-axis on the right. The two front clumps are
    projected on to each other in the right sequence of figures.}
  \label{fig:M112std_proj}
\end{figure*}

We now investigate a more dynamical situation (model 18), in
which two clumps are pushed past a third which was initially shadowed,
and are wrapped around into the tail, forming a pillar.  This was how
the short-lived `pillar A' formed in model 16, and
it also crudely models the later evolution of pillar B.  In this
scenario, the three massive clumps are placed at positions
$[0.45,0.72,0.6]$, $[0.45, 0.72, 0.9]$, and $[0.75, 0.72, 0.72]$,
giving us two clumps in front shadowing the third clump, all in the
plane $y=0.72\,$pc. This plane also contains the source at
$[-2.0,0.72,0.72]$.  The properties of the two front clumps were
modified to have masses $M=27.7\,\msun$, peak overdensities of 250,
and scale radii of $0.1125\,$pc, while the shadowed clump has mass of
$M=28.3\,\msun$, overdensity of 500, and scale radius of $0.09\,$pc,
as in the previous model.  The formation of a pillar in this
configuration is quite sensitive to the relative clump positions; if
the front clumps are too close together or too dense they do not move
past the shadowed one without disrupting it and if they are too far apart
they never merge with the shadowed clump to form a single clump/pillar
configuration.

We first ran this simulation using cooling model C1 and a volume
rendering of the results is shown in Fig.~\ref{fig:M112std_Vol}.  The
pillar develops more slowly than in the previous scenario, taking
around $200-250\,$kyr to form, because the gas has to travel further
to get into the tail.  Initially the front two clumps are compressed
and slowly accelerated, seen after $50\,$kyr in
Fig.~\ref{fig:M112std_Vol}.  They collide obliquely with the shadowed
third clump, which then starts to accelerate due to a combination of
being exposed to ionising radiation and to the shocks driven through
it by the passage of the first two clumps.  Material from the remains
of the first two clumps sweeps into the tail region after $200\,$kyr,
producing the pillar-like structure seen in the last panel at
$250\,$kyr.  In this model the ET structure is formed dynamically
rather than being left over from initial conditions, and the column of
neutral gas resembles a pillar only after $250\,$kyr, unlike the
previous model.

As one might expect from the initial conditions, this column is quite
asymmetric; it is much broader in one direction than the other at late
times.  The volume rendering shows it from the narrower perspective;
projections at 20\degr\ from perpendicular to the pillar are shown in
Fig.~\ref{fig:M112std_proj} with edge-on projections at right and
face-on at left.  The gas in the trunk is at volume densities of
$1-3\times10^4\pcmc$, but projecting through the narrow axis does not
build up enough column density to resemble a pillar whereas the
projection through the broad axis clearly resembles an ET at
$250\,$kyr.

The velocity profiles show a broad range of velocities which vary
significantly with perspective and over the evolution of the system.
We do not see a strong signature here as was seen in
Fig.~\ref{fig:M101std_proj}.  One interesting feature is seen in the
right $P-V$ diagram at $150\,$kyr.  This shows high velocity gas from
the front two clumps which has pushed past the third and moved into
the trunk away from the head.  This is a similar profile to that
of~\citet{LefLaz94} for gas streaming from the head into the tail of a
cometary globule.  The feature is not clearly seen in the left-hand panel
because the tail is strongly re-expanding along the LOS, leading to a
broad velocity profile which has the opposite slope from
head to tail compared to the right-hand panel.  Here transverse motions are
once again masking the true trend in recession velocity from the star
along the pillar.

\section{Effect of cooling on pillar formation}
\label{sec:cooling}

\begin{figure*}
  \centering 
  \includegraphics[width=1.0\textwidth]{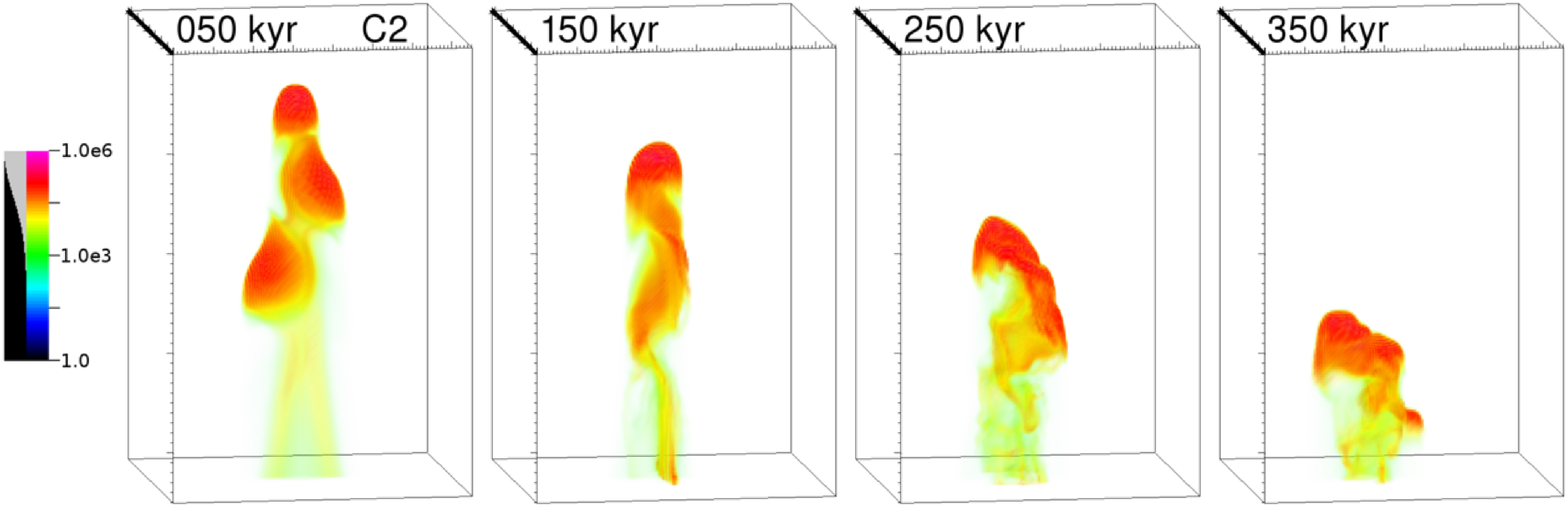}
  \caption{As Fig.~\ref{fig:M101std_Vol}, but with the simulation run
    with cooling model C2.  Simulation times are indicated on each
    panel.}
  \label{fig:M101c15_Vol}
\end{figure*}

\begin{figure*}
  \centering 
  \includegraphics[width=1.0\textwidth]{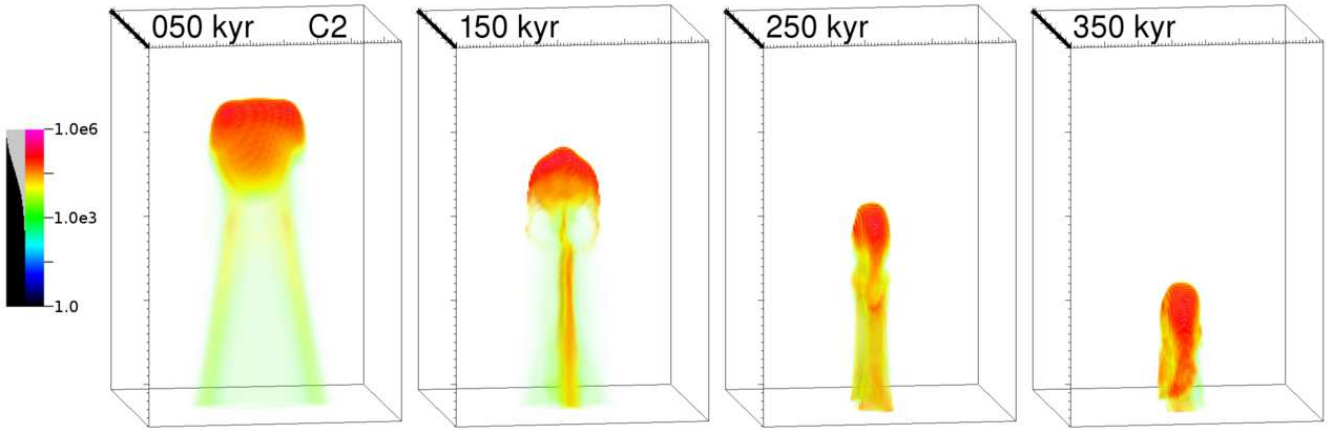}
  \caption{As Fig.~\ref{fig:M112std_Vol}, but with the simulation run
    with cooling model C2, and with frames at different times.}
  \label{fig:M112c15_Vol}
\end{figure*}

\begin{figure*}
  \centering 
  \includegraphics[width=0.83\textwidth]{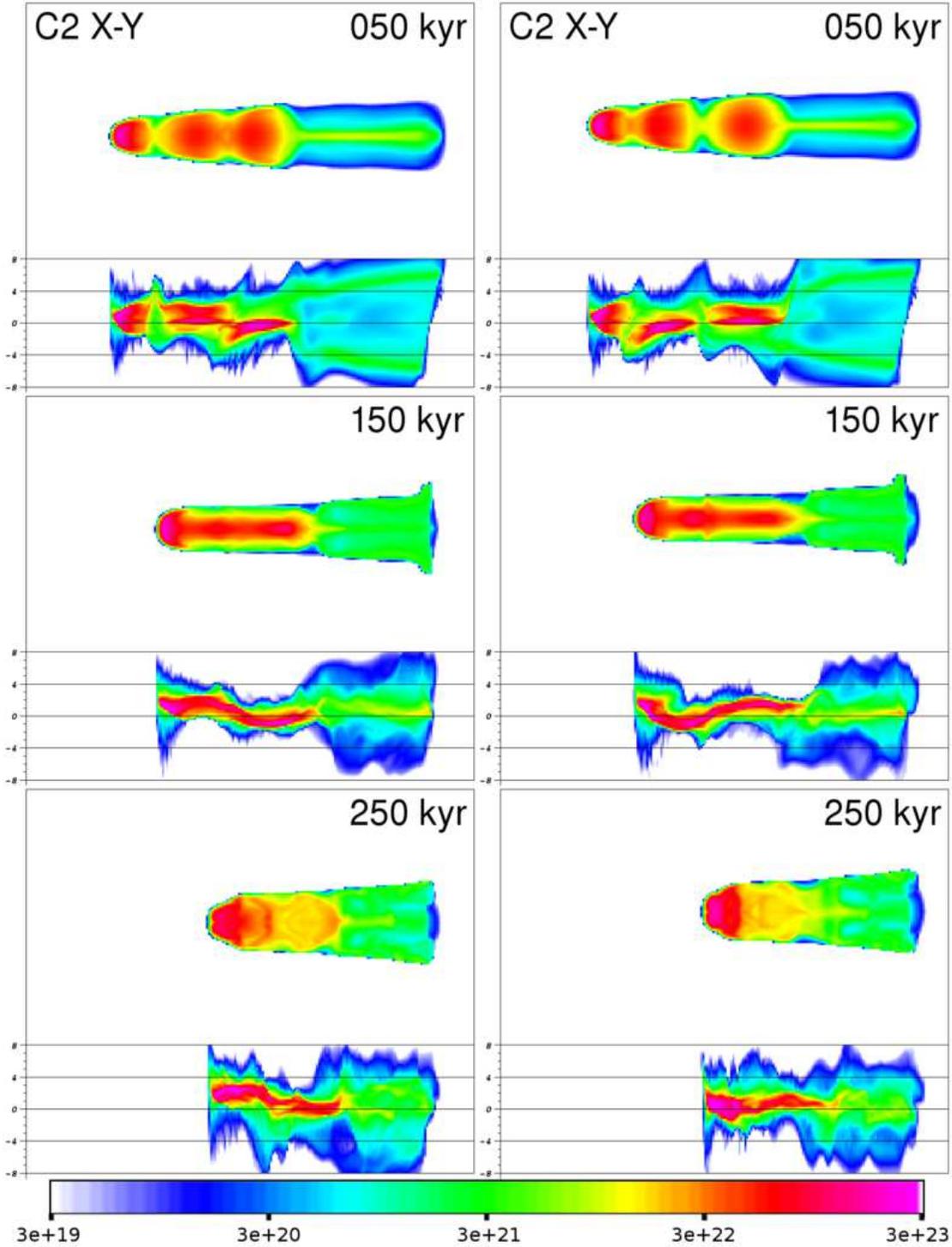}
  \caption{Projections through the ``co-linear clumps'' 3D simulation
    shown in Fig.~\ref{fig:M101c15_Vol} with cooling model C2.  Only
    the projection at 20\degr\ to the $z$-axis is shown, but from the
    front of the domain at left and from the rear at right.  This
    highlights the effect transverse gas motions can have on velocity
    profiles.}
  \label{fig:M101c15_proj}
\end{figure*}

\begin{figure*}
  \centering
  \includegraphics[width=0.83\textwidth]{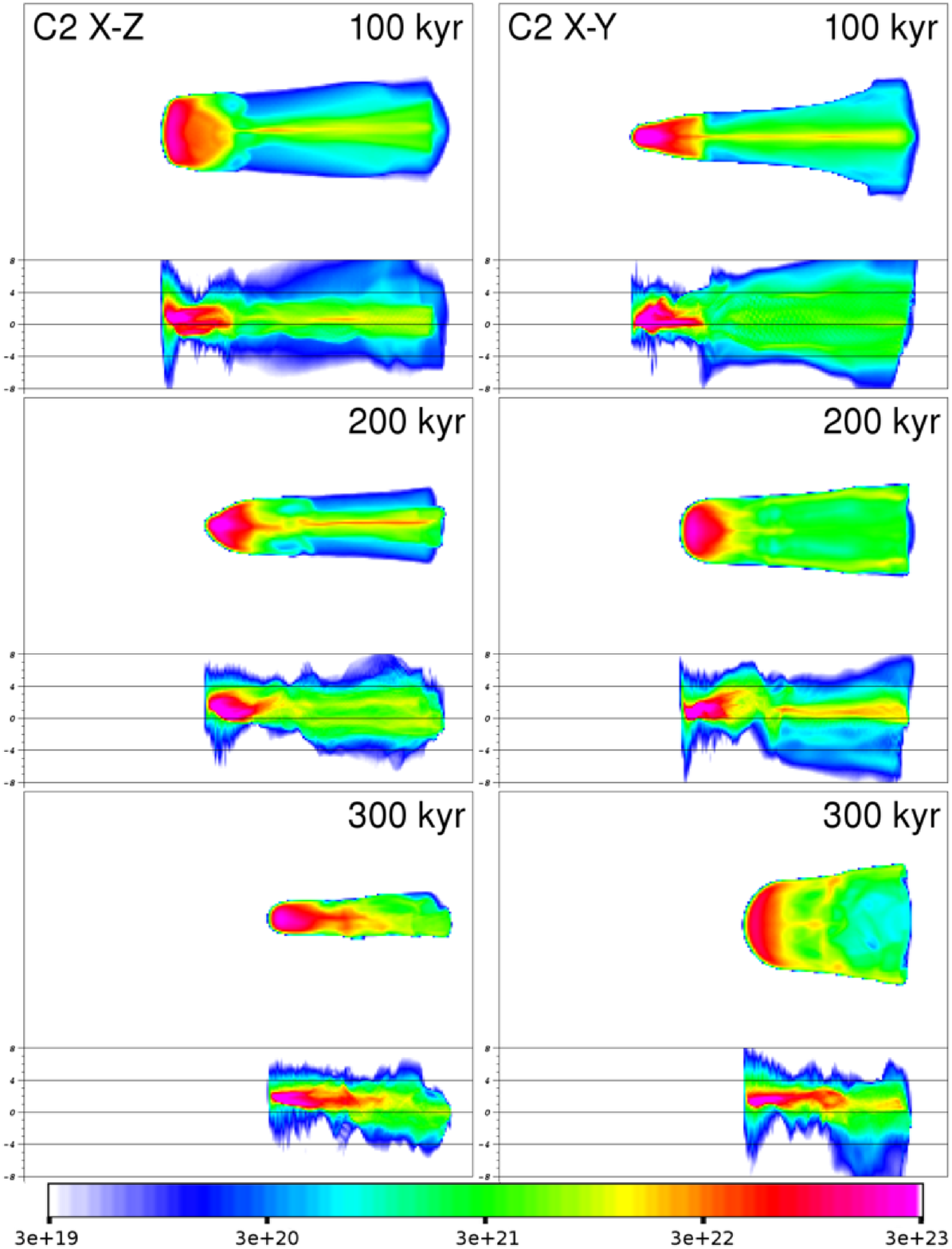}
  \caption{Projections through the ``triangle of clumps'' 3D
    simulation shown in Fig.~\ref{fig:M112c15_Vol} with the C2 cooling
    model.  The projections are at 20\degr\ to the $y$-axis on the
    left, and to the $z$-axis on the right, as in
    Fig.~\ref{fig:M112std_proj} but shown at different times to
    reflect the slower evolution of this model.}
  \label{fig:M112c15_proj}
\end{figure*}

\citet{MelArtHenEA06} use a cooling model similar to our C1 model,
with relatively little neutral gas cooling, to model the global
expansion of an \hii\ region into a turbulent density field.
Individual features are not highly resolved, but it is clear that
their results (e.g.\ their fig.~5) are qualitatively similar to ours,
showing fairly smooth coherent pillar-like structures.  In particular
both they and we do not find I-front instabilities developing, or any
fragmentation which is a feature of simulations with strong neutral
cooling.  This is undoubtedly due to the thermal physics, since it has
been shown by~\citet{Wil02} and~\citet{WhaNor08}, among others, that
instabilities in D-type I-fronts are strongest when the neutral gas
can cool rapidly.  These instabilities can generate dense finger-like
structures~\citep[e.g.][]{WhaNor08,MizKanPouEA06} but it is unclear if
they could generate something as massive and large as the M16 complex
of pillars.

Other
authors~\citep[e.g.][]{LefLaz94,WilWarWhi01,LorRagEsq09,GriNaaWalEA09}
have used a thermal model where the neutral gas is cold and
isothermal, leading to strong fragmentation and instability in the
I-front.  Observationally,~\citet{WhiNelHolEA99} showed that the M16
pillars have cold cores ($T\sim10-20\,$K), warm trunks or fingers
($T\sim60\,$K), and a `hot' shell of surrounding gas at
$T\sim250-320\,$K.  To study the possible effects of these temperature
variations,~\citet{MiaWhiNelEA06} use a much more detailed thermal
model to study the head of a pillar numerically, but at the expense of
using very low resolution for the dynamics.  More
recently,~\citet{HenArtDeCEA09} have fit a number of functions to
detailed calculations of cooling and heating rates for their MHD
models of photo-evaporating clumps.  They found the shadowed neutral
gas was not isothermal, but had a range of temperatures comparable to
observed values.  These results show that while the isothermal
approximation is expected to be more realistic than our C1 model, it
is unclear whether it is sufficient to capture the details of ET
formation.

To investigate the influence neutral gas cooling has on our results,
we have repeated the simulations in the previous section using cooling
model C2, designed to be intermediate between C1 and the isothermal
approximation.  In comparison to~\citet{HenArtDeCEA09}, their
`molecular' cooling rate (which they denote $L_{\mathrm{PDR}}$) scales
as $L\propto\rho^{1.6}\,\mathrm{erg}\,\pcmcps$ whereas our C2 model
scales as $L\propto \rho$.  So while our model has significant neutral
gas cooling down to $100\,$K, very dense gas cools more rapidly in
their simulations, closer to an isothermal model.

Volume rendering of the linear and triangular clump configurations
with C2 cooling are shown in Figs.~\ref{fig:M101c15_Vol}
and~\ref{fig:M112c15_Vol}.  The evolution is slower than with C1 so we
show the results at times 50, 150, 250, and $350\,$kyr on the same
neutral gas number density scale.  Comparing the $50\,$kyr panels
between the C1 and C2 runs, the main change is immediately apparent.
The extra neutral gas cooling makes shocks more compressive and slower
moving.  This can be clearly seen in the shadowed region, where the
converging shock is about $10\times$ denser in
Fig.~\ref{fig:M101c15_Vol} than in Fig.~\ref{fig:M101std_Vol}.  The
cooling also affects the exposed clump; the radiatively driven
implosion phase is slower and the clump is compressed to a greater
degree.  The rocket effect is less effective on this exposed clump:
firstly there is a smaller surface area to absorb the photon flux;
secondly, the denser gas in the I-front and in the photo-evaporation
flow leads to faster recombinations and thus more photons are required
to ionise an atom and keep it ionised until it flows away from the
I-front.

Comparing the $150\,$kyr and $250\,$kyr panels of
Figs.~\ref{fig:M101std_Vol} and~\ref{fig:M101c15_Vol}, the difference
is dramatic.  The leading clump has fully merged into the second in
the C1 run, and the implosion phase is followed by a strong
re-expansion.  Re-expansion in the C2 run is much weaker because most
of the heat generated is radiated away before it can instigate a
rebound.  We obtain a much narrower and denser structure which
resembles an ET for at least $200\,$kyr.  The initial configuration is
enhanced rather than disrupted by the photo-ionisation process.

Fig.~\ref{fig:M112c15_Vol} shows a very similar story for the triangle
of clumps.  The system's evolution is slower, the shocks are more
compressive, and there is less re-expansion after the implosion phase.
The general picture from these two models is that ETs are longer,
narrower, denser, longer-lived, and have more substructure with the C2
cooling prescription.  While qualitatively the same evolutionary
scenario plays out, quantitatively the neutral gas cooling has a
significant effect on the results.

Projections through the C2 runs are shown in
Figs.~\ref{fig:M101c15_proj} and~\ref{fig:M112c15_proj}.
Fig.~\ref{fig:M101c15_proj} shows the projection at 20\degr\ as in the
left sequence of Fig.~\ref{fig:M101std_proj} for C1, but the two
sequences show projections from the front and rear perspectives.  The
column density maps are similar from both perspectives and it is clear
that the structure resembles an ET at all times shown.  The left $P-V$
diagrams show the same trend as in the C1 model, where gas at the head
is receding more rapidly from the star than the trunk is, but this
trend takes significantly longer to become established.  The right
hand $P-V$ diagrams also show the head receding rapidly, but the gas
from the partially shadowed clumps has very different LOS velocities
due to transverse motions.  This particular case shows how difficult
it is to infer anything about a formation scenario from a $P-V$
diagram.  With non-axisymmetric initial conditions, transverse motions
mix with the recession velocity in an unpredictable way and are a
strong contaminant when one wants to measure the recession velocity
along the length of a pillar.  Fig.~\ref{fig:M112c15_proj} shows
projections face-on and edge-on through the triangle of clumps
scenario, again at 20\degr\ to perpendicular.  The same asymmetry as
shown in Fig.~\ref{fig:M112std_proj} is apparent, but to a lesser
degree, and the C2 run shows that the column of gas resembles an ET in
both directions from $200-300\,$kyr, albeit much more convincingly in
the left hand sequence.  The $P-V$ diagrams show very little trend in
velocity along the length of the pillar at most times, but again it
can be seen in the $200\,$kyr panels that there is high velocity gas
moving from the head into the tail region, as was seen in the C1
model.

\section{Discussion}
\label{sec:discussion}

\citet{WilWarWhi01} performed axisymmetric simulations with a similar
aim to our work -- to investigate the mechanisms by which ETs can
form.  They modelled parallel rays rather than treating the radiation
from a source at a finite distance.  In most of their models,
\citet{WilWarWhi01} start with a dense ($n\sim10^4\pcmc$) layer of gas
at the boundary most distant from the star.  This served both to stall
the I-front and to provide a reservoir of dense gas from which to
build up a pillar.  This is a very similar picture to that modelled
more recently in 3D by~\citet{RagHenVasEA09}, who studied small
($l\sim0.03\,$pc) columns of dense gas forming in a photo-ionised
region. Their initial conditions had an effectively infinite reservoir
of neutral gas shadowed by smaller clumps. Evaporating gas from the
reservoir flowed into the shadowed regions forming dense columns as it
recombined and cooled.  These authors have demonstrated a mechanism by
which ETs can form given a sufficiently large reservoir of dense gas
behind a shadowing clump.  We consider alternate scenarios, however,
where the ET must be built up from clumps of comparable mass and from
low density inter-clump gas.  We find that this is sufficient to
produce pillars when the gas is in certain configurations, but we also
find that they do not form as readily as found by~\citet{WilWarWhi01}.
This is quite important: while ETs are seen in many \hii\ regions,
structures resembling heaps and corrugations which are not elongated are
much more common. The mechanisms by which ETs form cannot therefore be
too efficient or many more ETs would be observed.

The fact that most \hii\ regions seem to have one or two ET structures
led~\citet{WilWarWhi01} to suggest that they may be long-lived
objects, with lifetimes comparable to the \hii\ region, a proposition
supported by their simulation results.  An alternative scenario is
that they are short lived ($\sim100\,$kyr), but multiple generations
of them occur during the expansion of an \hii\ region.  For models
where the gas is not already organized in a linear structure, we do
not find any pillars forming in less than $150\,$kyr.  Adding extra
cooling in the neutral gas only lengthens this formation time.
Additionally, it takes about $250-300\,$kyr before more massive and
long-lived pillars start to form dynamically.  This is the rough
time-scale for dense neutral gas to be accelerated, pushed past other
dense clumps, and stretched out into a long tail. We suggest it would
be difficult to dynamically generate a parsec-scale dense ET from
static initial conditions in less than $150\,$kyr, and our results are
consistent with the claim by~\citet{WilWarWhi01} that ETs are likely
to be long-lived structures.

\begin{figure}
  \centering 
  \includegraphics[width=0.48\textwidth]{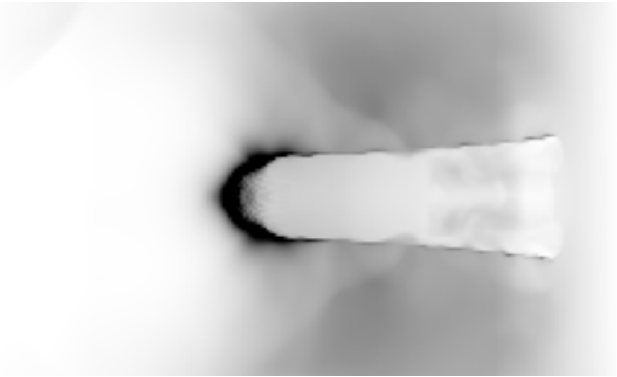}
  \caption{Simulated H$_{\alpha}$ emission map from our simulation of
    a pillar forming from three almost co-linear clumps using cooling
    model C2.  The map can be compared to projected density in the
    bottom left panel of Fig.~\ref{fig:M101c15_proj}.  The scale used
    is linear with the strongest emission in black and deliberately
    saturated to bring out low contrast features.}
  \label{fig:Halpha_map}
\end{figure}

\subsection{Gas Pressure}
Observational age constraints are thus far not very stringent.  An
upper age limit for ETs is set by the age of their \hii\ region.  The
free-fall time of the densest parts of the pillars does not constrain
their age as the gas is being actively compressed and so the free-fall
time is changing as the structures evolve.  If the pressure difference
between the neutral and ionised gas is very large, however, this may
indicate that the dense gas has not had time to dynamically respond to
the photo-ionisation, and so the structures must be young, perhaps
younger than their sound crossing time.  In M16, the gas pressure at
the base of the ionised photo-evaporation flow has been estimated
by~\citet{HesScoSanEA96} to be
$p_g/k_B=6\times10^7\,\mathrm{cm}^{-3}\,\mathrm{K}$.  The internal
pressure in the densest parts of the pillars was estimated
by~\citet{WhiNelHolEA99} to be
$p_g/k_B=3.5\times10^7\,\mathrm{cm}^{-3}\,\mathrm{K}$.  These values
are based on temperature and density estimates in ionised and neutral
gas respectively.  \citet{WhiNelHolEA99} interpreted this pressure
difference as suggesting the ETs in M16 may be young, with the dense
clumps currently undergoing the early stages of radiation-driven
implosion~\citep{Ber89}.  In our simulations the peak pressure at the
ionisation front varies between $p_g/k_B=5\times10^7$ and
$2\times10^8\,\mathrm{cm}^{-3}\,\mathrm{K}$, and in the neutral ETs
between $p_g/k_B=2\times10^6$ and
$5\times10^7\,\mathrm{cm}^{-3}\,\mathrm{K}$.  These values are
consistent with the observations, but we found that an equilibrium
state was never reached in our models.  The pressure varies by large
factors both in time and in space along the length of the ET, and the
peak pressure changes in time according to the instantaneous density
at the I-front.  These results suggest that the observed pressure
difference (a factor of $1.7$) should not be taken as strong evidence
that the pillars are young since we find pressure variations much
larger than this within the dense neutral gas.  A caveat to this is
that our thermal modelling of the neutral gas is crude and gas
pressure is sensitive to this, so it would require more detailed
modelling to make a definitive statement.  We can say that simulations
with both cooling models C1 and C2 have these pressure variations, and
we believe the dynamical nature of the ETs which form in our models
will always generate significant pressure gradients within the ETs.
It would be interesting to compare this with structures formed via
I-front instabilities to see how their internal dynamics differ.

\subsection{Morphology of ETs}
Molecular emission traces the projected mass density of ETs and has
shown them to be rather more clumpy than they appear in optical
data~\citep{Pou98,WhiNelHolEA99}, with significant density variations
along their length.  The heads of the pillars are the densest regions,
with clumpy lower density gas found in the trunk.  Our simulations
support this clumpy model for the underlying density structure; in
some of our models the clumps maintain their identity for hundreds of
kyr (e.g.~Fig.~\ref{fig:M101c15_proj}).  We also find that the highest
density gas is always found in the head of the pillars because the
strongest compression is always found ahead of the I-front which
drives the dynamics.  We have also calculated optical emission maps
due to recombination radiation (e.g.~H$_{\alpha}$) in the manner
descibed by~\citet{HenArtGar05}, using a constant dust opacity as
in~\citet{MelArtHenEA06}.  An image from the co-linear clumps
simulation (model 17) at $250\,$kyr is shown in
Fig.~\ref{fig:Halpha_map}; this can be compared to the bottom left
panel of Fig.~\ref{fig:M101c15_proj} which shows projected neutral gas
density and is closer to what we expect from a molecular emission map.
It is clear that the `optical' image shows a much smoother structure
which appears more like the HST images of M16.  In particular the
dense parts of the pillar's trunk are completely dark in the optical
image while there is substructure in the projected density.  Our image
has less substructure than the HST images; this is likely due to the
limited spatial resolution of our simulations, and possibly because
our cooling models do not promote the development of I-front
instabilities.

\subsection{Position-Velocity Diagrams}
The full 3D nature of our simulations has allowed us to calculate
line-of-sight velocity profiles through the ETs which include
asymmetric transverse motions.  We have shown that these motions can
largely determine the observed profile (Fig.~\ref{fig:M101c15_proj}).
This has implications for interpreting the $P-V$ diagrams
in~\citet{Pou98} and~\citet{WhiNelHolEA99}, in particular for the
largest M16 pillar which has a velocity gradient that changes sign
from the head to the tail.  This was interpreted as evidence that the
head and tail are separate structures, but this is difficult to
reconcile with optical observations.  Our results suggest a resolution
to this apparent contradiction: this pillar could have started out as
two separate dense clumps which have since merged but have kept their
identity in velocity-space.  Fig.~\ref{fig:M101c15_proj} shows just
such a case where the clumps gain opposing transverse velocities,
giving the projected velocity profile an $S$-shape which is maintained
for $>100\,$kyr despite the clumps having merged into a pillar much
earlier.

For the smallest of the M16 pillars, \citet{Pou98} notes that the
shadowing implies it is closer to us than the ionising stars so we
expect the tail to stream towards us faster than the head.  In fact
the opposite is true.  Our results also offer two possible
explanations for these observations: (A) it could be due to a strong
transverse motion in the pillar if it is seen close to edge-on, or (B)
it could be formed from a column of dense gas closely aligned with the
radiaton propagation direction.  Our models of this situation show
that for at least some of the pillar's evolution the head can be
receding from the star faster than the tail.

\subsection{Influence of other physical processes}
We have already discussed in detail the effects of neutral gas cooling
on our results, but there are a number of other processes which could
contribute significantly.  Since massive stars generate strong winds,
\citet*{RagSteGon05} studied photo-evaporating clumps interacting with
a stellar wind, finding that when the photo-ionisation was
sufficiently intense the photo-evaporation flow effectively shielded
the clumps from the wind.  This gives us confidence that ionising
radiation is the main driver of ET formation.  \citet{RagHenVasEA09}
studied photo-ionisation models with a basic treatment of diffuse
radiation.  They found there was no strong difference between runs
with the diffuse radiation and runs where they used the on-the-spot
approximation.  This suggests that, while diffuse radiation would
undoubtedly change our results somewhat, it is unlikely to make a
dramatic difference.  \citet{EsqRag07} studied the effects of
self-gravity on the photo-ionisation of a dense cloud of gas using a
two-temperature equation of state (i.e.\ the neutral gas is
isothermal) which gives rise to strong instabilities and fragmentation
associated with the photo-ionisation process.  Interestingly, they
found that self-gravity had little effect on the overall process of
the evaporation and fragmentation of the massive clump, and was only
significant in determining the properties of the densest sub-clumps
produced by the fragmentation.  This supports our implicit assertion
in this work that pressure forces are the dominant driver in the
evolution of photo-evaporating clumps, at least in the early stages.

\section{Conclusions}
\label{sec:conclusions}
Using moderately high resolution 3D radiation-hydrodynamics
simulations of clumpy density fields exposed to ionising radiation
from a point source, we have investigated how effectively shadowing
can generate pillar-like structures.  We have shown that even in the
absence of self-gravity or I-front instabilities, large parsec-scale
ETs can form dynamically solely due to this shadowing.

Of our 2D and 3D models, about 20 per cent produced long-lived
pillars.  This is more due to our choices of initial conditions than
how easy or difficult it is to generate ETs.  Nevertheless we did find
certain density fields more conducive than others to forming pillars.
The most successful of these contain both large low density regions
where the I-front can propagate far from the source, and also massive
clumps with sufficiently high density and inertia to stall the I-front
for $>100\,$kyr.  Pillars are formed with diameters comparable to
those of the clumps that give rise to them.

Simulations of specific configurations of massive clumps show that
pillar-like initial conditions evolve with a different velocity
signature to configurations where the ETs form dynamically from clumps
that are not initially co-linear.  For a simulation with three clumps
initially almost co-linear we find the head of the ET recedes from the
source more rapidly than the shadowed trunk which has not been exposed
to radiation.  For dynamically forming ETs, gas streams into the
shadowed trunk past the head, and is thus moving faster than, or at a
comparable speed to, the pillar's head. This could produce a
noticeable observational signature from which these two formation
mechanisms can be distinguished, but we find that the $P-V$ diagrams
vary significantly with viewing angle due to transverse gas motions.
This variation increases with the degree of asymmetry and the
transverse motions significantly contaminate attempts to measure
recession velocity in the pillars.  These transverse motions offer an
explanation for the features seen in $P-V$ diagrams for the pillars in
M16.

We have shown that neutral gas cooling has a very strong influence on
the modelling results.  Stronger cooling produces ETs which take
longer to form dynamically, are narrower and denser, and are more
resistant to the rocket effect and hence live longer.  This shows that
the complex chemistry and thermal physics in molecular clouds may play
a crucial role in ET formation and evolution.  For the specific case
of almost co-linear clumps, the initial pillar-like configuration was
slowly disrupted in simulations with the C1 cooling model.  By
contrast it was enhanced with the C2 cooling model, also proving to be
long-lived.

We have not addressed magnetic fields in this work.
\citet{HenArtDeCEA09} have studied the photo-ionisation of a
magnetised globule, finding that strong uniform fields can have a
significant influence on the evolution of the photo-ionisation
process.  In future work we will investigate the effects of more
realistic thermal physics, as well as the presence of magnetic fields,
on the results we have presented here.

\section*{Acknowledgments}
JM is funded by the Irish Research Council for Science, Engineering
and Technology: funded by the National Development Plan.  AJL is
funded by a Schr\"{o}dinger Fellowship from the Dublin Institute for
Advanced Studies.  Figures were generated using the \textit{VisIt}
visualisation tool.  The authors wish to acknowledge the SFI/HEA Irish
Centre for High-End Computing (ICHEC) for the provision of
computational facilities and support, and are grateful to Turlough
Downes and Garrelt Mellema for suggestions which improved the
presentation of this work.  We thank the referee for very helpful
comments which significantly improved the paper.

\bibliography{refs}

\end{document}